\documentclass{article}
\usepackage{PRIMEarxiv}
\usepackage[utf8]{inputenc} 
\usepackage[T1]{fontenc}    
\usepackage{enumitem}
\setlist{nosep}
\usepackage{hyperref}       
\usepackage{url}            
\usepackage{booktabs}       
\usepackage{titlesec}
\titlespacing{\section}{0pc}{0pc}{0pc}
\titlespacing{\subsection}{0pc}{0pc}{0pc}
\usepackage{float}
\usepackage{cite}
\usepackage{amsmath,amssymb,amsfonts}

\usepackage[monochrome]{xcolor}

\usepackage{cite}
\usepackage{hyperref}
\usepackage{graphicx}
\usepackage{algpseudocode}
\usepackage{graphicx}
\usepackage{textcomp}
\usepackage{xcolor}
\usepackage{subfigure}
\usepackage{multirow}
\usepackage{tabularx}
\usepackage{gensymb}
\usepackage{nidanfloat}
\usepackage{svg}
\usepackage{optidef}
\usepackage{diagbox}
\usepackage{threeparttable}
\usepackage{placeins}
\usepackage{float}
\usepackage{appendix}
\usepackage[ruled,vlined]{algorithm2e}
\newcommand{\linebreakand}{%
  \end{@IEEEauthorhalign}
  \hfill\mbox{}\par
  \mbox{}\hfill\begin{@IEEEauthorhalign}
}
\usepackage{nomencl}
\usepackage{etoolbox}
\def\BibTeX{{\rm B\kern-.05em{\sc i\kern-.025em b}\kern-.08em
    T\kern-.1667em\lower.7ex\hbox{E}\kern-.125emX}}
    
\renewcommand\nomgroup[1]{%
  \item[\bfseries
  \ifstrequal{#1}{A}{Superscripts}{%
  \ifstrequal{#1}{B}{Parameters}{%
  \ifstrequal{#1}{C}{Sets}{%
  \ifstrequal{#1}{D}{Indices}{%
  \ifstrequal{#1}{E}{Input Data}{%
  \ifstrequal{#1}{F}{Decision Variables}{%
  \ifstrequal{#1}{G}{Functions}{}}}}}}}%
]}

\makenomenclature
\begin{document}


\title{Toward Efficient Transportation Electrification of Heavy-Duty Trucks: Joint Scheduling of Truck Routing and Charging\\
}


\author{
  Mikhail A. Bragin, Zuzhao Ye, Nanpeng Yu \\
  Department of Electrical and Computer Engineering \\
  University of California \\
  Riverside, CA, USA\\
  \texttt{mbragin@engr.ucr.edu, zye066@ucr.edu, nyu@ece.ucr.edu} \\
  }

\maketitle

\thispagestyle{fancy}

\begin{abstract}
The timely transportation of goods to customers is an essential component of economic activities. However, heavy-duty diesel trucks used for goods delivery significantly contribute to greenhouse gas (GHG) emissions within many large metropolitan areas, including Los Angeles, New York, and San Francisco. To reduce GHG emissions by facilitating freight electrification, this paper proposes Joint Routing and Charging (JRC) scheduling for electric trucks. The objective of the associated optimization problem is to minimize the cost of transportation, charging, and tardiness. A large number of possible combinations of road segments as well as a large number of combinations of charging decisions and charging durations 
leads to a combinatorial explosion in the possible decisions electric trucks can make. The resulting mixed-integer linear programming (MILP) problem is thus extremely challenging because of the combinatorial complexity even in the deterministic case. Therefore, a Surrogate Level-Based Lagrangian Relaxation (SLBLR) method is employed to decompose the overall problem into significantly less complex truck subproblems. In the coordination aspect, each truck subproblem is solved independently of other subproblems based on 
the values of Lagrangian multipliers.
In addition to serving as a means of guiding and coordinating trucks, multipliers can also serve as a basis for transparent and explanatory decision-making by trucks. Testing results demonstrate that even small instances cannot be solved using the off-the-shelf solver CPLEX after several days of solving. The SLBLR method, on the other hand, can obtain near-optimal solutions within a few minutes for small cases, and within 30 minutes for large ones. Furthermore, it has been demonstrated that as battery capacity increases, the total cost decreases significantly; moreover, as the charging power increases, the number of trucks required decreases as well. 

\end{abstract}

\keywords{Heavy-Duty Vehicles \and Transportation Electrification \and Surrogate Level-Based Lagrangian Relaxation}

\section{Introduction}

The transportation of goods to customers is an essential economic activity. An effective logistical design is critical in ensuring the delivery of goods on time. However, as a result of the excessive use of transportation in large metropolitan areas, such as Los Angeles, New York, and San Francisco, many of these areas are heavily congested and polluted. In addition to public transportation, commercial transportation, such as large freight trucks that deliver goods, contributes significantly to greenhouse gas emissions. More than 40\% of greenhouse gas emissions in California are attributed to the transportation sector, as reported in \cite{California}. Transportation has been widely acknowledged for its environmental impacts (such as climate change due to emissions) \cite{Savelsbergh201650th, pantelidis2022node, guillet2022electric, Fernandez2022arc, abdelwahed2020evaluating}. Although heavy-duty trucks make up only 1\% of all vehicles, their emissions account for 25\% of all vehicle emissions \cite{TruckEmissions}. In light of this, electrification of the transportation sector, especially heavy-duty vehicles, offers a promising means of reducing greenhouse gas emissions \cite{asadi2022monotone, guillet2022electric, cao2022joint, kabir2021joint, zhang2020deep} as well as addressing fossil fuel shortages \cite{zhang2020deep}. 

Despite significant advances in electrifying a range of vehicles, such as passenger cars and transit buses, the availability of electric heavy-duty trucks in the market remains notably limited. There are several reasons for this disparity. First, the energy consumption per mile for heavy-duty trucks is significantly higher than that of passenger cars, requiring larger batteries and consequently, driving up initial costs. For example, while a passenger car uses roughly 0.3 kWh per mile, a heavy-duty truck consumes more than 2 kWh \cite{weiss2020energy}. Second, compared to transit buses, heavy-duty trucks are driven considerably farther each year (62,000 miles compared to 43,000 miles, according to \cite{MilesDriven}). This increased operational range necessitates the fleets of electric heavy-duty trucks to opt for more frequent, higher-powered, or longer-duration charging. Lastly, the charging infrastructure for heavy-duty trucks is largely underdeveloped. The construction of charging stations, requiring high power capacity and substantial space, is likely to be expensive, leading to a limited number of such facilities in the near future. Each station, consequently, is expected to face high demand, thereby leading to potentially long wait times and elevated risks of delivery delays. Given these challenges, it is crucial to develop strategically tailored algorithms for joint routing and charging as a key means for ensuring sustainable operations of heavy-duty trucks upon their electrification.

Existing literature extensively addresses various aspects of electric vehicle (EV) operations. The majority of the research has been conducted on passenger cars and transit buses. The operations of electric passenger cars (or light-duty vehicles) usually fall into the category of electric vehicle routing problems (EVRP) \cite{kucukoglu2021electric, erdelic2019survey}. This field focuses on finding optimal routes and charging plans for electric vehicles. Current EVRP research can be generally distinguished by its objective functions and corresponding constraints. The common objective function typically includes operational cost factors such as travel distance \cite{jia2021bilevel, goeke2019granular}, travel time \cite{karakativc2021optimizing}, charging costs \cite{li2015multiple}, energy consumption \cite{penna2016hybrid, barco2017optimal
}, and the mix of two or more factors \cite{li2015multiple, zhao2019heuristic, bac2021optimization, setak2019mathematical, sassi2014vehicle}. The compulsory constraint is the battery energy level constraint, while common constraints include pickup-and-delivery \cite{goeke2019granular, ghobadi2021multi} and vehicle loading capacity \cite{jia2021bilevel,lu2019bi, bac2021optimization}. Beyond the common considerations, \cite{trivino2019joint} proposed routing EVs in a power network for vehicle-to-grid (V2G) applications, thus optimizing the timing and location of energy trading. Autonomous EVs, due to their increased flexibility without drivers, hold even more potential for V2G applications \cite{james2017autonomous, iacobucci2019optimization}. Ref. \cite{sassi2017electric} considered the operations of mixed fleets of electric and conventional vehicles, which is particularly meaningful during the transition phase to EVs. Enabling additional customer visits by walking or other means of transportation while the EV is being charged could potentially reduce EV downtime, as suggested by \cite{cortes2019electric}. There are also studies targeting individual EVs instead of the fleet. Ref. \cite{lu2019integrated} studied the routing of individual EVs under an improved Dijkstra algorithm and with forecasting of charging cost. More comprehensive literature reviews can be found in \cite{afroditi2014electric, mukherjee2014review, erdelic2019survey, kucukoglu2021electric}. In a short summary, despite variations in problem settings, the fundamental concerns of the EVRP remain consistent.

The operations of electric transit buses are generally categorized as electric bus scheduling problems (EBSP). In contrast to EVRP, EBSP tends not to emphasize routing since bus routes are typically predetermined. Instead, EBSP concentrates on optimal bus assignment to cover the entire transit system timetable \cite{perumal2022electric,alwesabi2020electric}. The objective usually encompasses the costs of vehicles and associated charging infrastructure (as the transit system is usually self-contained) \cite{rogge2018electric, wei2018optimizing}, in addition to the operational costs considered in EVRP. Attention is often given to charging station capacity due to the significant size of transit buses and the substantial costs related to high-power charging \cite{foda2023generic, ye2022decarbonizing}. There are many interesting works in EBSP. \cite{alwesabi2020electric} considers the installation of dynamic wireless charging lanes such that the battery size and the station charging time can both be reduced. \cite{li2021routing} considers treating electric buses as temporary mobile power sources to provide grid services. Battery degradation is also considered by multiple studies \cite{wang2020optimal,zhou2022electric}, as replacing the high-capacity battery of transit buses can be costly.

The operation of electric heavy-duty trucks, while sharing commonalities with passenger cars and transit buses, also exhibits unique characteristics: 
\begin{itemize}
    \item First, the tasks of heavy-duty trucks vary daily and lack fixed timetables, unlike transit buses. This necessitates joint routing and charging scheduling, positioning this problem as an enhanced variant of EVRP rather than EBSP. 
    \item Second, truck operations share a common concern with EBSP regarding charging station capacity due to their substantial dimensions. At any given time, the number of vehicles being charged at a station cannot exceed the number of available chargers. To accommodate this requirement, an additional time dimension is introduced, which is common in EBSP \cite{foda2023generic, ye2022decarbonizing}, but not found with EVRP. Specifically, the time horizon needs to be discretized to enable the modeling of charging station capacity limit in any time period. This additional dimension amplifies the model's complexity.
    \item Last but not least, the operation of heavy-duty trucks carries its own unique considerations. These include treating time constraints as tardiness instead of compulsory time windows and modeling the influence of cargo on the energy consumption rate - it necessitates the introduction of additional binary variables.
\end{itemize}

As a result, the operation of heavy-duty trucks demands a specialized model formulation, distinct from those of transit buses or passenger cars. While the existing literature on EVRP has also considered certain unique features of heavy-duty trucks (\cite{setak2019mathematical} considers charging station capacity and \cite{kancharla2020electric} considers the impacts of cargo on the energy consumption rate), to the best of our knowledge, no comprehensive study has been found that encompasses all these aspects. Past studies specifically targeting heavy-duty trucks have exhibited limitations. \cite{pelletier2018charge} sought to minimize electric freight truck charging costs but assumed fixed routes, optimizing only the charging schedule without considering route alterations. \cite{zhao2021vehicle} employed a bi-level heuristic for joint charging and routing of electric trucks but overlooked the essential factor of charging station capacity. In summary, there is a pressing need for a comprehensive and efficient framework for the joint charging and routing of electric heavy-duty trucks.

Regarding solution methods, EBSPs are generally solved exactly by commercial solvers \cite{alwesabi2020electric}, as they naturally skip the complexity added by routing on road networks. EVRP, on the other hand, can only be solved exactly for small instances due to their additional routing requirements. As such, EVRP generally relies on heuristic or meta-heuristic methods for larger instances \cite{kucukoglu2021electric}. Given the wide adoption of meta-heuristic methods, we will provide a brief overview. Commonly used meta-heuristic methods include genetic algorithms \cite{abdulaal2016solving, karakativc2021optimizing,shao2018electric}, variable neighborhood search \cite{wang2020time,bac2021optimization,kancharla2020electric,lu2019bi}, ant colony optimization \cite{jia2021bilevel,meng2020route}, iterated local search \cite{cortes2019electric,sassi2014vehicle,penna2016hybrid}, tabu search \cite{lin2021electric,goeke2019granular,li2015multiple}, simulated annealing \cite{ghobadi2021multi,setak2019mathematical,kuccukouglu2019hybrid}, large neighborhood search \cite{keskin2021simulation,zhao2019heuristic,schiffer2018designing}, and differential evolution algorithms \cite{barco2017optimal}. While the above methods represent a broad range of solution techniques, the list is by no means exhaustive. A common element among these methods is the use of heuristics to reduce the solution time of EVRP which features high combinatorial complexity. However, due to their heuristic nature, these algorithms cannot be systematically improved based on theory. Moreover, these meta-heuristics largely rely on case-by-case developed, hand-crafted heuristics, meaning they are not always adaptable to other variants. In contrast, although Lagrangian Relaxation methods may also require heuristics, the solutions obtained can be continuously improved. This improvement is achieved through efficient coordination using Lagrangian multipliers.
Lagrangian Relaxation was indeed considered by \cite{james2017autonomous, zhang2022smart, lin2021electric} to reduce the complexity of EVRP. However, their final solution method did not consider the coordination of Lagrangian multipliers; instead, they reverted to the use of traditional meta-heuristics. As such, the benefit of Lagrangian Relaxation is not fully utilized.

We present an overview of the typical characteristics of the operational problems for various types of vehicles in Table \ref{table-literature}, based on our literature review. Note that Table \ref{table-literature} does not aim to provide an exhaustive summary. Instead, it seeks to highlight common features across similar types of vehicles.

\begin{table}[!h]
\caption{Compare operational problems of different types of vehicles in the literature.} 
\centering 
\begin{threeparttable}
\begin{tabular}{c l c c c} 
\toprule 
Vehicle type &  & Passenger car & Transit bus  &  Heavy-duty truck  \\
\toprule 
\multirow{4}{*}{Model feature} & Objective function$^{(1)}$	&	Op	&	Op + Veh + Infra	&	Op	\\
& Routing	& $\textcolor{blue}{\checkmark}$	& &	$\textcolor{blue}{\checkmark}$ \\
& Charging scheduling & $\textcolor{blue}{\checkmark}$	& $\textcolor{blue}{\checkmark}$	& $\textcolor{blue}{\checkmark}$\\
& Fleet or individual & Both	& Fleet	& Fleet\\
\midrule
\multirow{4}{*}{Constraints} & Time$^{(2)}$ & TW & TB & TD \\
& Energy & $\textcolor{blue}{\checkmark}$ & $\textcolor{blue}{\checkmark}$ & $\textcolor{blue}{\checkmark}$	\\
& Pickup-delivery & $\textcolor{blue}{\checkmark}$ & & $\textcolor{blue}{\checkmark}$ \\
& Station capacity &	&  $\textcolor{blue}{\checkmark}$ & $\textcolor{blue}{\checkmark}$\\
& Cargo$^{(3)}$ &	& &	$\textcolor{blue}{\checkmark}$ \\
\midrule
\multirow{3}{*}{Solution method} & & Heuristic & \multirow{3}{*}{Exact} & \multirow{3}{*}{Meta-heuristic} \\
 & & Meta-heuristic & Exact & Meta-heuristic \\
 & & Exact & & \\
\bottomrule 
\end{tabular}
\begin{tablenotes}
\footnotesize
\item (1) Op: Operational costs, e.g. travel distance/time and charging costs; Veh: Vehicle procurement cost; Infra: Charging infrastructure cost.
\item (2) TW: Time window; TB: Trip-based constraint, i.e. the arrival time of a trip should be ahead of its connecting trip; TD: Tardiness or soft time window, appearing as a penalty term.
\item (3) The loading/unloading of cargo will create two levels of energy consumption rates that require additional binary constraints to handle.
\end{tablenotes}
\end{threeparttable}
\label{table-literature} 
\end{table}

The rest of the paper is organized in the following manner. We present a novel formulation of a Joint heavy-duty vehicle fleet Routing and Charging (JRC) problem in Section \ref{problemformulation}. Compared with EBSP, this formulation introduces additional routing requirements; Compared with general EVRP, this formulation adds a completely new time dimension to decision variables to enable the modeling of charging station capacity limit, similar to that of EBSP. The loading/unloading of cargo and their impacts on the energy consumption rate are also modeled through a set of additional binary variables. We consider the new formulation as an enhanced variant of EVRP.


To effectively manage the increased complexity introduced by the innovative problem formulation and to address the limitations of existing methodologies, we utilize the principles behind the Surrogate ``Level-Based'' Lagrangian Relaxation (SLBLR) approach \cite{Bragin22}, the nuances of which are elaborated in Section \ref{SolutionMethodology}. The SLBLR possesses several key desirable properties:
\begin{enumerate}
\item It drastically reduces the combinatorial complexity via decomposition into truck subproblems; the method essentially ``reverses'' the inherent combinatorial complexity of the underlying problem;

\item The surrogate component of the methodology allows for bypassing the need for achieving optimal solutions for the truck subproblems at each iteration as long as the ``surrogate optimality condition'' is satisfied. This feature, which is lacking in the standard Lagrangian Relaxation appears to be the dominant as will be demonstrated in Section \ref{NumericalTesting};

\item The ``level-based'' coordination nature of our method exploits the linear convergence rate (fastest possible in the dual space). While other methods \cite{goffin1998convergence} also utilize the ``level-based'' ideas, their level adjustment is heuristic in nature, while our adjustment is informed and decision-based leading to a faster adjustment of level values thus allowing us to exploit linear convergence potential more efficiently; 

\item Penalty terms are used for constraint violations to accelerate convergence\footnote{Here convergence can be understood both in dual and, perhaps, more importantly in the primal space, since the penalties allow to obtain feasible solutions.} and are even capable of ``pushing'' the total constraint violation to zero thereby avoiding the
need for coding other heuristics for primal solution recovery. 


\end{enumerate}

Besides the above properties, the SLBLR method inherently benefits from the following features: 
\begin{enumerate}
\item \textbf{Flexibility and Adaptability.} The SLBLR method exhibits model-agnostic capabilities. In essence, the same algorithm can be easily adapted to different model formulations, ensuring its versatility and adaptability across a wide array of problem scenarios;

\item \textbf{Continuous Improvement.} The SLBLR approach 
enables a continuous refinement of solutions. Unlike meta-heuristic methods employed for EVRP, the SLBLR technique
facilitates continual improvements of solution achieved through the efficient coordination of Lagrangian multipliers;

\item \textbf{Economic Viability and Explainability.} The method's \textit{price-based feature} stems from the economic theory, wherein Lagrangian multipliers are interpreted as ``shadow prices'' that adjust according to the ``supply and demand'' principle, ensuring that the obtained solutions are economically viable.
In terms of the JRC problem, the ``demand'' is the number of trucks needing to charge at a particular node of the road network, and the ``supply'' is the number of chargers. When the ``demand'' exceeds the ``supply'' (charging stations are oversubscribed), ``prices'' (not to be confused with the charging price) increase thereby discouraging trucks from making less ``economically viable'' decisions, e.g., from driving to an oversubscribed charging station thereby losing time and facing a higher penalty for late shipments.
Multipliers thus possess the intuitive explainability feature behind the underlying decision making such as the grounds for the re-routing of trucks along the longer paths, which lead to lower overall costs. The same logic can be applied to the actual demand and supply of goods to be transported;

\item \textbf{Smooth Convergence.} The SLBLR method (like other LR-based methods) is an iterative approach and optimal multipliers are obtained by taking a series of steps along multiplier-updating directions. While traditionally the subgradient directions were used, surrogate subgradients were shown to be more beneficial for saving computational effort and leading to smoother convergence. The choice of stepsizes is not only crucial for guaranteeing convergence but also important for achieving a high rate of convergence, e.g., fast \textit{iteration-wise} convergence. The  \textit{level-based feature} of the method rests upon the fundamental principles Polyak stepsize \cite{Polyak69}; moreover, our method resolves the issue of the unavailability of knowledge about the optimal value required for fast convergence within \cite{Polyak69}. 
\end{enumerate}
Recently, although within a different problem context, SLBLR \cite{Bragin22} demonstrated fast convergence, and several discrete programming problems have been solved while beating the default CPLEX by several orders of magnitude in terms of CPU time. 

A series of case studies in Section \ref{NumericalTesting} demonstrate that fast coordination of truck subproblems through the SLBLR method yields near-optimal feasible solutions for problems with different sizes: from five trucks to fifty. In addition, a financial impact analysis is conducted, demonstrating that increased capacity of truck batteries leads to a significant reduction in costs; meanwhile, increased charger power also results in fewer trucks being required to transport goods.  

\section{Problem Description and Formulation}\label{problemformulation}


\noindent \textbf{Description.} Consider a transportation network with
\begin{enumerate} 
\item A set of nodes $n \in \mathcal{N}$;
\item A set of road segments $r \in \mathcal{R}$ with each segment characterized by a ``starting node'' $s(r)$ and ``ending node'' $e(r)$;\footnote{Note that starting and ending nodes are not unique for a road segment since the road segments are generally bi-directional.}
\item A set of heavy-duty electric vehicles (trucks) $v \in \mathcal{V}$;
\item A set of trips $t \in \mathcal{T} = \{1,\dots, T\}$ with each trip being a one-way trip either from a depot at node $n^{depot} \in \mathcal{N}^{depot} \subset \mathcal{N}$ to a port $n^{port} \in \mathcal{N}^{port} \left(\subset \mathcal{N}\right)$ or from $n^{port}$ to $n^{depot};$
\item A set of nodes $n^{chrg} \in \mathcal{N}^{chrg} \left(\subset \mathcal{N}\right)$ containing charging stations;
\item A lookahead horizon $p \in \mathcal{P} = \{1,2, \dots, P\}$ with $P$ being the total number of time periods, and 
\item Products $pr_n \in \mathcal{PR}$ that needs to be delivered either from depot to port $n = n^{port}$ or from port to depot $n = n^{depot}$. 
\end{enumerate} 

The time required to travel through a road segment $r$
is $T^{{trvl}}_{r,p}$, which may depend on the time of the day $p$ as well as the direction traveled (e.g., for any two adjacent nodes $n_1$ and $n_2$, generally, $T^{{trvl}}_{(n_1,n_2),p} \neq T^{{trvl}}_{(n_2,n_1),p}$); the charging cost is $C_{p,n^{chrg}}$ per every time period $p$ at node $n^{chrg}$; and the labor driving cost per time period is $C^{lbr}$. 


In this paper, the following assumptions are made:
\begin{enumerate}
    \item Each truck $v$ is designated to a certain depot $n^{depot}_{v}$ and 
    a certain port $n^{port}_v$.\footnote{The difference between $n^{depot}$ and $n^{depot}_v$ is that $n^{depot}$ is a dummy index from a set $\mathcal{N}^{depot}$ whereas $n^{depot}_v$ is a specific depot node where truck $v$ is located.} Whether cargo needs to be delivered from a depot to a port or vice versa is determined through optimization as explained ahead; 
    \item Each truck is expected to be fully charged overnight, i.e., the initial charge is 100\%. This assumption is not strict, since the methodology developed further will be able to handle any level of the initial charge, if feasible; 
    \item Each port is a node $n^{port} \in \mathcal{N}^{port}$ whereby cargo is loaded, and each port is equipped with a charging station; 
    \item Each depot is a node $n^{depot} \in \mathcal{N}^{depot}$ whereby the cargo is unloaded, and each depot is also equipped with a charging station;
    \item Other nodes of the network $n \in \mathcal{N}\setminus \{\mathcal{N}^{port}\cup \mathcal{N}^{depot}\}$ may or may not contain a charging station; 
    \item Driving along a road segment $r$ is \textit{non-preemptive}, that is, once started driving at time $p$, the truck will arrive at the end of the road segment after $T^{trvl}_{r,p}$ time periods; 
    \item Each truck $v$ can take several trips (not necessarily $T$) during a scheduling horizon as long as the last trip ends at $n^{depot}_v$; 
    \item The increase in the truck battery's state of charge is a linear function of charging time; 
\end{enumerate}

\begin{figure}[H]
  \centering
    \includegraphics[trim=0 0 0 0,  width=0.8\linewidth, scale=0.5]{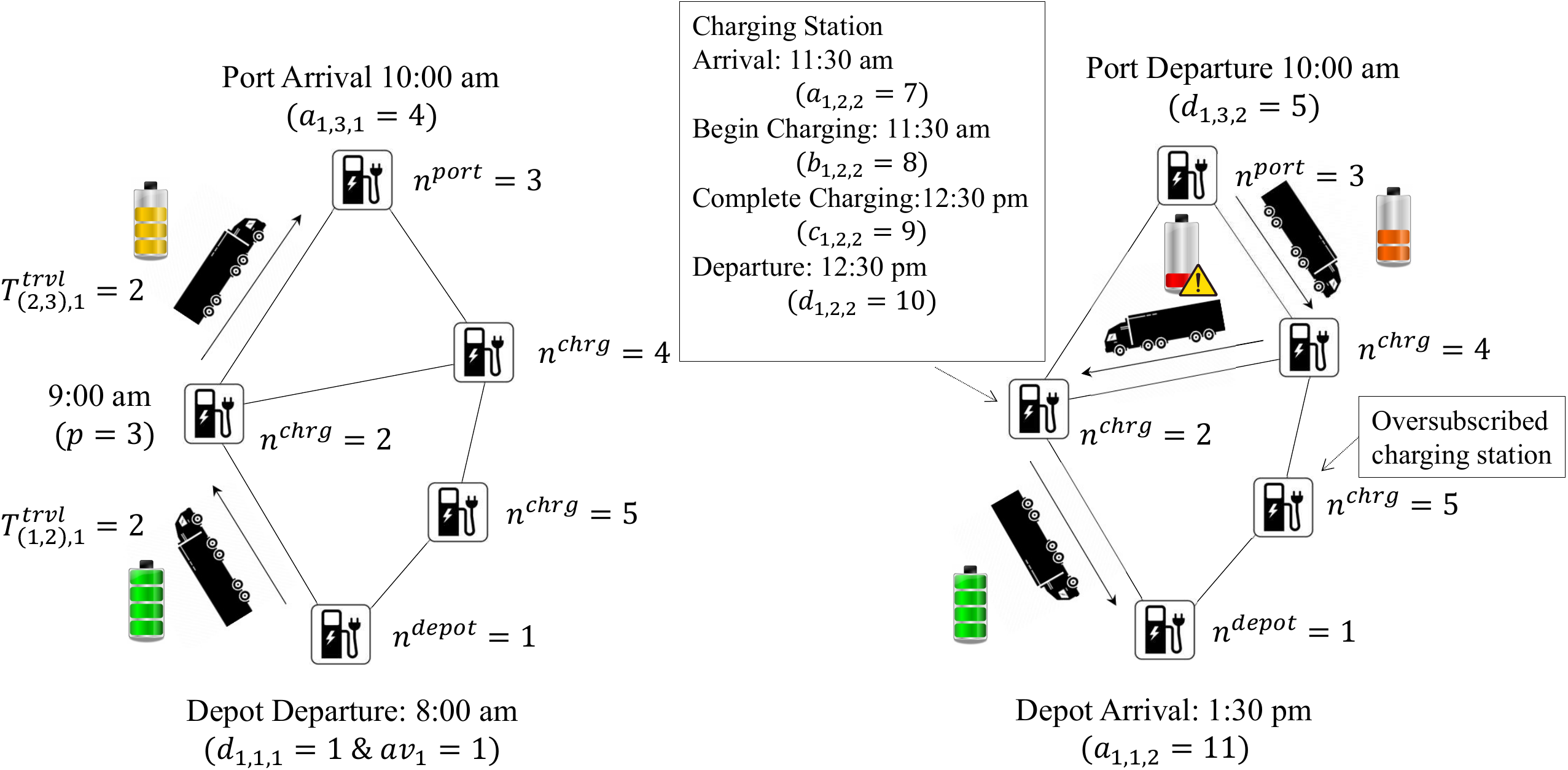}
    \caption{A potential routing and charging scenario for a sample electric truck.}
    \label{fig1}
\end{figure}

Figure \ref{fig1} schematically illustrates one possible routing-charging scenario of one truck (truck 1). After overnight charging at the depot (charging station located at node $n^{depot}_1 = 1$), the truck sets out to pick up the cargo at the port. Having a sufficient amount of charge, the truck drives past the charging station at node $n^{chrg}=2$. At node $n^{port}_1=3,$ the truck loads (here the loading, as well as unloading times, are assumed negligible). After loading, the truck chooses road segment $\left(3,4\right)$.
At node $n^{chrg}=4,$ the truck needs to decide whether to charge at $n^{chrg}=2$ or $n^{chrg}=5.$ In this example, it is assumed that the charging station $n^{chrg}=5$ is oversubscribed leading to a potentially long waiting time. The truck then travels to node $n^{chrg}=2$ to charge. After charging, the truck returns to the depot. 

The goal is to transport all goods from the depot nodes to the port nodes (and vice versa) while co-minimizing the total charging and labor costs, as well as the tardiness penalties due to goods shipped late.
To achieve the above goal, the truck needs to decide when and which road segment to take, which charging station to choose, and for how long. The cargo picked up at the port needs to be unloaded at the depot. The above process needs to be repeated until the demand at the depot is satisfied. In the same fashion, the demand at the port is satisfied. These considerations will be mathematically formulated next.

The rest of the section is organized as follows. In subsection \ref{tripwise}, constraints for scheduling the movement as well as the charging of a heavy-duty truck within one-way trips are modeled. Loading/unloading constraints to connect one-way trips are introduced in subsection \ref{tripconnecting}. The multiple-truck scenario will then be considered in subsection \ref{multipletruck} by introducing charging station ``capacity'' constraints. 
\vspace{-1mm}
\subsection{Trip-Wise Constraints}\label{tripwise}
\vspace{-1mm}
To capture trucks traveling and charging within a one-way trip,
several sets of binary decision variables are introduced: 
\begin{enumerate} 
\item To capture the beginning of a trip, let $x^{trvl,dprtr}_{v,n,p,t} = 1$ if truck $v$ chooses to depart from node $n$ at time $p$ at trip $t$ 
and $x^{trvl,dprtr}_{v,n,p,t} = 0$ otherwise; 
\item The ``arrival'' binary decision variable $x^{trvl,arrvl}_{v,n,p,t}$ is similarly defined; 
\item To capture the status of charging, let $x^{chrg}_{v,n^{chrg},p,t} = 1$ if truck $v$ chooses to charge at node $n^{chrg}$ during time period $p$ and $x^{chrg}_{v,n^{chrg},p,t} = 0$ otherwise; 
\item To capture beginning and completion of charging, let $x^{chrg,bgn}_{v,n^{chrg},p_1,t} = 1$ and $x^{chrg,cmplt}_{v,n^{chrg},p_2,t} = 1$ if truck $v$ began charging during trip $t$ at node $n^{chrg}$ at time $p_1$ and ended charging at time $p_2,$ respectively. 
\end{enumerate}


\noindent \textbf{Truck Availability.} For every truck $v$, the first trip $t=1$ starts at a depot $n = n^{{depot}}_v$ after the time $av_v$ truck $v$ is available. To capture this condition (as well as several conditions thereafter), binary variable $x^{trip}_{v,t}$ is introduced: 
\begin{flalign}
& x^{trip}_{v,1} = 1 \Rightarrow d_{v,n^{depot}_v,1} \geq av_v, \forall \left(v \in \mathcal{V}\right),   \label{eq1}
\end{flalign}
where $d_{v,n_v^{depot},1}$ is the departure time. 

\noindent \textbf{Departure-Arrival Relationship.} To make sure that truck $v$ can only depart from at most one node during every trip $t$, the following set of constraints is introduced: 
\begin{flalign}
& \sum_{n \in  \mathcal{N}} x^{trvl,dprtr}_{v,n,p,t} \leq 1, \forall \left(v \in \mathcal{V}, p \in \mathcal{P}, t \in \mathcal{T}\right).    \label{eq3}
\end{flalign}
Moreover, truck $v$ can only depart once, therefore, summation over the time periods is also required:  
\begin{flalign}
& \sum_{p \in  \mathcal{P}} x^{trvl,dprtr}_{v,n,p,t} \leq 1, \forall \left(v \in \mathcal{V}, n \in \mathcal{N}, t \in \mathcal{T}\right).    \label{eq4}
\end{flalign}
The corresponding departure time $d_{v,n,t}$ is determined through the following set of constraints: 
\begin{flalign}
& \sum_{p \in \mathcal{P}} p \cdot x^{trvl,dprtr}_{v,n,p,t} = d_{v,n,t}, \forall \left(v \in \mathcal{V}, n \in \mathcal{N}, t \in \mathcal{T}\right). \label{eq5}
\end{flalign}

The above constraints \eqref{eq3}-\eqref{eq5} also hold for the binary arrival indicator $x^{trvl,arrvl}_{v,n,p,t}$ and integer arrival time $a_{v,n,t}$ decision variables and for brevity are not shown. 

Once truck $v$ departs from node $s(r)$ (starting node of road segment $r$), it needs to arrive at one of the nodes $e(r)$ (ending node of road segment $r$) $T^{trvl}_{r,p}$ units of time later (due to non-preemptiveness), which is captured through the following set of constraints: 
\begin{flalign}
& x^{trvl,dprtr}_{v,n,p,t} = 1 \Rightarrow \sum_{r \in  \mathcal{R}: s(r) = n} x^{trvl,arrvl}_{v,e(r),p+T^{trvl}_{r,p}-1,t} = 1, \label{eq6} \forall \left(v \in \mathcal{V}, n \in \mathcal{N}, t \in \mathcal{T}, p \in \mathcal{P}\right).    
\end{flalign}

The above relationship goes both ways: if truck $v$ arrived at node $n$, it must have departed in the past $T^{trvl}_{r,p}$ units of time ago from one of the nodes directly connected by road segments with node $n$: 
\begin{flalign}
& x^{trvl,arrvl}_{v,n,p,t} = 1 \Rightarrow \sum_{r \in  \mathcal{R}: e(r) = n} x^{trvl,dprtr}_{v,s(r),p-T^{trvl}_{r,p}+1,t} = 1,  \forall \left(v \in \mathcal{V}, n \in \mathcal{N}, t \in \mathcal{T}, p \in \mathcal{P}\right).  \label{eq7}
\end{flalign}
In the above constraints, one unit of time is added or subtracted from the travel time as appropriate because of the discrete nature of the time periods: departure is understood at the top of the time period while arrival is understood at the end of the time period. For example, if the truck departed at the beginning of time 3 and traveled 3 units of time, then the arrival will be at 3 + 3 - 1 = 5, that is, the truck travels during three time periods 3, 4, and 5.  

Moreover, once truck $v$ arrives at node $n$, it needs to depart from $n$ in the future unless truck $v$ reaches the destination $\in \{n^{port}_v \cup n^{depot}_v\}$.\footnote{Note that after arriving at the destination, the truck may still depart. However, the departure will happen during the next trip $t+1,$ which will be discussed in subsection \ref{tripconnecting}.} The following sets of constraints, capture this relationship: 
\begin{flalign}
& x^{trvl,arrvl}_{v,n,p,t} = 1 \Rightarrow \sum_{p' \in  \mathcal{P}: p' \geq p + 1} x^{trvl,dprtr}_{v,n,p,t} = 1, \label{eq8} \forall \left(v \in \mathcal{V}, n \in \mathcal{N}\setminus\{n^{port}_v \cup n^{depot}_v\}, t \in \mathcal{T}, p \in \mathcal{P}\right).    
\end{flalign}




\noindent \textbf{Charging and Discharging.} While travelling, truck $v$ discharges at rate $\Delta s^{dchrg,ldd} \%$ per time period, if loaded, and with rate $\Delta s^{dchrg,empty} \%$, if empty. To capture the effect of the above discharge rates, 
binary decision variables $x^{ld}_{v,t}$ are introduced to indicate whether truck $v$ is loaded during trip $t$ ($x^{ld}_{v,t} = 1$) or empty ($x^{ld}_{v,t} = 0$). With the help of the above-defined ``departure'' and ``arrival'' decision variables, the state of charge becomes: 
\begin{flalign}
& x^{trvl,dprtr}_{v,n,p,t}  =  1 \; \wedge \;  x^{trvl,arrvl}_{v,e(r),p+T^{trvl}_{r,p}-1,t} = 1 \; \wedge   x^{ld}_{v,t} = 1 \Rightarrow \nonumber \\  & s^{chrg}_{v,e(r),t} = s^{chrg}_{v,n,t} - \frac{\Delta s^{dchrg,ldd}_{r}}{100} \cdot T^{trvl}_{r,p} \cdot x^{trvl,arrvl}_{v,e(r),p+T^{trvl}_{r,p}-1,t}, \label{eq9}
\\ & \forall \left(v \in \mathcal{V}, n \in \mathcal{N}, r \in \mathcal{R}: s(r) = n, t \in \mathcal{T}, p \in \mathcal{P}: p+T^{trvl}_{r,p}-1 \leq P\right),  \nonumber 
\end{flalign}
where $s^{chrg}_{v,n,t} \in [0,1]$ is a continuous decision variable capturing the state of charge for the battery of truck $v$ at node $n$ during trip $t$; $\wedge$ denotes the conjunction (also referred to as the ``logical AND''). Correspondingly, if truck $v$ is empty during trip $t$ then the above constraint becomes:  
\begin{flalign}
& x^{trvl,dprtr}_{v,n,p,t}  =  1 \; \wedge \;  x^{trvl,arrvl}_{v,e(r),p+T^{trvl}_{r,p}-1,t} = 1 \; \wedge  x^{ld}_{v,t} = 0 \Rightarrow \nonumber \\& s^{chrg}_{v,e(r),t} = s^{chrg}_{v,n,t} - \frac{\Delta s^{dchrg,empty}_{r}}{100} \cdot T^{trvl}_{r,p} \cdot x^{trvl,arrvl}_{v,e(r),p+T^{trvl}_{r,p}-1,t}, \label{eq10}
\\ & \forall \left(v \in \mathcal{V}, n \in \mathcal{N}, r \in \mathcal{R}: s(r) = n, t \in \mathcal{T}, p \in \mathcal{P}: p+T^{trvl}_{r,p}-1 \leq P\right).  \nonumber 
\end{flalign}

If truck $v$ arrives at node $n^{chrg}$ equipped with chargers, it can make a decision to charge; such a decision is captured through a binary variable $x^{chrg}_{v,n^{chrg},p,t}$. Assuming that $\Delta s^{chrg} \%$ is the charge rate per time period, \eqref{eq9} is modified as: 
\begin{flalign}
& x^{trvl,dprtr}_{v,n,p,t}  =  1  \wedge   x^{trvl,arrvl}_{v,e(r),p+T^{trvl}_{r,p}-1,t} = 1  \wedge   x^{ld}_{v,t} = 1 \Rightarrow \nonumber \\&s^{chrg}_{v,e(r),t} = s^{chrg}_{v,n,t} - \frac{\Delta s^{dchrg,ldd}_{r}}{100} \cdot T^{trvl}_{r,p} \cdot x^{trvl,arrvl}_{v,e(r),p+T^{trvl}_{r,p}-1,t} + \label{eq11}
 \sum_{p' \in \mathcal{P}} \frac{\Delta s^{chrg}_{r}}{100} \cdot x^{chrg}_{v,e(r),p',t}, \\ &\nonumber  \forall \left(v \in \mathcal{V}, n \in \mathcal{N}, r \in \mathcal{R}: s(r) = n, t \in \mathcal{T}, p \in \mathcal{P}: p+T^{trvl}_{r,p}-1 \leq P\right).  
\end{flalign}
In the above equation $e(r)$ needs to be a member of a set $\mathcal{N}^{chrg}$. Equation \eqref{eq10} is modified in the same way. 

In \eqref{eq11}, the amount of energy charged is determined through the number of time periods used for charging. However, the appropriate timing for charging needs to be determined to make sure that electric truck $v$ can charge only if it arrived at and departed from node $n^{chrg}$. Therefore, the following constraints are introduced: 
\begin{flalign}
& \sum_{p' = 1}^{p - 1} x^{trvl,arrvl}_{v,n^{chrg},p',t} \geq x^{chrg}_{v,n^{chrg},p,t}, \forall \left(v \in \mathcal{V}, n^{chrg} \in \mathcal{N}^{chrg}, t \in \mathcal{T}, p \in \mathcal{P}\right).   \label{eq12}
\end{flalign}
In the equation above, the upper limit of summation $p-1$ is once again used because of the discrete nature of the time periods; if truck $v$ arrives at the end of the time period $p-1$, truck $v$ can start charging at time $p$. 

The charging will not be possible after the departure, which is captured through the following set of constraints:   
\begin{flalign}
& 1 - \sum_{p' = 1}^{p} x^{trvl,drptr}_{v,n^{chrg},p',t} \geq x^{chrg}_{v,n^{chrg},p,t}, \forall \left(v \in \mathcal{V}, n^{chrg} \in \mathcal{N}^{chrg}, t \in \mathcal{T}, p \in \mathcal{P}\right).   \label{eq13}
\end{flalign} 

If truck $v$ departs at time $p$, it is naturally no longer eligible for charging starting at time $p$. 

The beginning $b_{v,n^{chrg},t}$ and completion $c_{v,n^{chrg},t}$ times of charging are captured by introducing binary variables 
$x^{chrg,bgn}_{v,n^{chrg},p,t}$ and $x^{chrg,cmplt}_{v,n^{chrg},p,t}$ in the same way as departure times are captured within \eqref{eq5}. The same relations as in \eqref{eq3} and \eqref{eq4} hold for $x^{chrg,bgn}_{v,n^{chrg},p,t}$ and $x^{chrg,cmplt}_{v,n^{chrg},p,t}$ as well. These binary variables are linked to $x^{chrg}_{v,n^{chrg},p,t}$ in the following ways: 
\begin{flalign}
& x^{chrg,bgn}_{v,n^{chrg},p,t} \geq x^{chrg}_{v,n^{chrg},p,t} - x^{chrg}_{v,n^{chrg},p-1,t}, \forall \left(v \in \mathcal{V}, n^{chrg} \in \mathcal{N}^{chrg}, t \in \mathcal{T}, p \in \mathcal{P}\setminus\{1\}\right),
\label{eq14}  
\end{flalign}
\begin{flalign}
& x^{chrg,cmplt}_{v,n^{chrg},p-1,t} \geq x^{{chrg}}_{v,n^{chrg},p-1,t} - x^{chrg}_{v,c,p,t}, \label{eq15}  \forall \left(v \in \mathcal{V}, n^{chrg} \in \mathcal{N}^{chrg}, t \in \mathcal{T}, p \in \mathcal{P}\setminus\{1\}\right).   
\end{flalign} 

\noindent \textbf{\textcolor{blue}{Unloading Start} Time Constraints.} Since the goal is to co-optimize charging cost and the tardiness, as stated at the beginning of the section, unloading start times\footnote{Here and throughout the rest of the paper, the ``unloading start times'' is used to indicate when unloading starts rather how long unloading lasts. As assumed earlier, the unloading duration is negligible.} will be captured through the use of the integer variables $u_{v,t,n,pr}$:
\begin{flalign}
& x^{ld}_{v,t} = 1 \Rightarrow u_{v,n,t,pr} = a_{v,n,t}, \label{eq16}  \forall \left(v \in \mathcal{V}, n \in \{\mathcal{N}^{port} \cup \mathcal{N}^{depot}\}, t \in \mathcal{T}, pr \in \mathcal{PR} \right).   
\end{flalign} 
If truck $v$ is loaded with product $pr$ during trip $t$ ($x^{ld}_{v,t} = 1$), it is implicitly assumed that the product will be \textcolor{blue}{unloaded} at a destination -- at the end of trip $t$, and the logical constraint \eqref{eq16} determines the \textcolor{blue}{unloading start} time at arrival. 
\vspace{-1mm}
\subsection{Trip-Connecting Constraints}\label{tripconnecting}
\vspace{-1mm}
The number of trips that need to be taken is decided through optimization. To capture whether a particular trip is taken, binary variables are introduced $x^{trip}_{v,t} = 1$, if truck $v$ takes trip $t$ and $x^{trip}_{v,t} = 0$, otherwise. If truck $v$ is to take the next trip $t+1$ (which is decided either at a depot $n^{depot}_v$ or at a port $n^{port}_v$), then the beginning time of the next trip ($t+1$) will relate to the arrival time as well as the charging completion time of the previous trip ($t$) in the following ways: 
\begin{flalign}
&  x^{trip}_{v,t+1} = 1 \; \wedge \; \sum_{p \in \mathcal{P}} x^{chrg}_{v,n,p,t} = 0 \Rightarrow d_{v,n,t+1} \geq a_{v,n,t} + 1, \forall \left(v \in \mathcal{V}, n \in \{n^{depot},n^{port}\}, t \in \mathcal{T}\setminus\{T\}\right),\label{eq17}
\end{flalign} 
\begin{flalign}
&  x^{trip}_{v,t+1} = 1 \;\wedge\; \sum_{p \in \mathcal{P}} x^{chrg}_{v,n,p,t} = 1 \Rightarrow d_{v,n,t+1} \geq c_{v,n,t} + 1 \label{eq18}
,\forall \left(v \in \mathcal{V}, n \in \{n^{depot},n^{port}\}, t \in \mathcal{T}\setminus\{T\}\right).
\end{flalign} 
The above constraints ensure that if truck $v$ is not charging (per \eqref{eq17}), then the next trip $t+1$ may start immediately after the completion of the previous trip $t$; if the truck needs to charge (per \eqref{eq18}), then the next trip $t+1$ may start immediately after the completion of the charging within previous trip $t$. 

Moreover, at a port ($n = n^{port}_v$), truck $v$ has to depart to return to a depot, regardless of whether the truck is loaded or not, which is captured through the following set of constraints: 
\begin{flalign}
&  x^{trip}_{v,t} = 1 \; \wedge \; \sum_{p \in \mathcal{P}} x^{chrg}_{v,n,p,t} = 0 \Rightarrow d_{v,n,t+1} \geq a_{v,n,t} + 1, \forall \left(v \in \mathcal{V}, n = n^{port}_v, t \in \mathcal{T}\right), \label{eq19}
\end{flalign} 
\begin{flalign}
&  x^{trip}_{v,t} = 1 \;\wedge\; \sum_{p \in \mathcal{P}} x^{chrg}_{v,n,p,t} = 1 \Rightarrow d_{v,n,t+1} \geq c_{v,n,t} + 1 \label{eq20}
, \forall \left(v \in \mathcal{V}, n = n^{port}_v, t \in \mathcal{T}\right). 
\end{flalign}

Since ports and depots are equipped with chargers, truck $v$ can make a decision to charge. In the next trip, the charging levels are determined as: 
\begin{flalign}
& x^{trip}_{v,t+1} = 1 \Rightarrow s^{chrg}_{v,n,t+1} = s^{chrg}_{v,n,t}
, \label{eq21}
\forall \left(v \in \mathcal{V}, n \in \{n^{depot}_v,n^{port}_v\}, t \in \mathcal{T}\setminus\{T\}\right),    
\end{flalign}
\begin{flalign}
& x^{trip}_{v,t} = 1 \Rightarrow s^{chrg}_{v,n,t+1} = s^{chrg}_{v,n,t}
, \label{eq22}
 \forall \left(v \in \mathcal{V}, n \in n^{{port}}_v, t \in \mathcal{T}\right).
\end{flalign}
Note that charging at nodes $\{n^{{depot}},n^{{port}}\}$ is already accounted for during trip $t$ within \eqref{eq9}-\eqref{eq11}. 

To ensure that a truck $v$ is not loaded without being scheduled for a trip, we utilize the binary variables $x^{ld}_{v,t}$, previously introduced in equations \eqref{eq9}-\eqref{eq11}. 
The following constraint set is introduced to maintain this logical consistency:
\begin{flalign}
& x^{{ld}}_{v,t} \leq  x^{trip}_{v,t}, \forall \left(v \in \mathcal{V},  t \in \mathcal{T}, pr \in \mathcal{PR}\right).  \label{eq23}  
\end{flalign}
Besides \eqref{eq9}-\eqref{eq11} and the above \eqref{eq23}, $x^{ld}_{v,t}$ will be used to capture the demand as explained ahead in subsection \ref{multipletruck}.

Lastly, to ensure the contiguity of trips, the following set of constraints is introduced: 
\begin{flalign}
& x^{trip}_{v,t+1} \leq x^{trip}_{v,t}, \forall \left(v \in \mathcal{V},  t \in \mathcal{T}\setminus \{T\}\right).  \label{eq24} 
\end{flalign}


\subsection{Truck-Coupling Constraints}\label{multipletruck}

\noindent \textbf{Demand Constraints.} Loaded trucks \textcolor{blue}{unload} cargo at the end of the trip. For inbound trips (to the depot), the \textcolor{blue}{unloaded} cargo $pr_{n^{depot}}$ will go towards satisfying the demand $D_{n^{depot},pr_{n^{depot}}}$ at the depot $n^{depot}$; for outbound trips (out from the depot), the \textcolor{blue}{unloaded} cargo $pr_{n^{port}}$ will go towards satisfying the demand $D_{n^{port},pr_{n^{port}}}$ at the port $n^{port}$:
\begin{flalign}
& \sum_{t \in \mathcal{T}^{inbnd} \subset \mathcal{T}, v \in \Omega_{\left(pr_{n^{port}},n_v^{depot}\right)}} x^{{ld}}_{v,t}  =  D_{n^{port}}, \forall(n^{port}  \in  \mathcal{N}^{port}, pr_{n^{port}}  \in  \mathcal{PR}), \label{eq25} \\
& \sum_{t \in \mathcal{T}^{otbnd} \subset \mathcal{T}, v \in \Omega_{\left(pr_{n^{depot}},n^{port}\right)}}   x^{{ld}}_{v,t}  =  D_{n^{depot}}, \forall(n^{depot}  \in  \mathcal{N}^{depot}, pr_{n^{depot}} \in \mathcal{PR}). \label{eq26}
\end{flalign}
Here $\mathcal{T}^{inbnd}$ is a subset of inbound trips (e.g., trips with odd numbers), and $\mathcal{T}^{otbnd}$ is a subset of outbound trips. 
Within \eqref{eq25}, the summation is over a set of trucks $\Omega_{\left(pr_{n^{port}},n^{depot}_v\right)}$ eligible to deliver product $pr_{n^{port}}$ from $n^{depot}_v$ to port $n^{port}$ toward satisfaction of demand $D_{n^{port}}$ if truck $v$ is loaded ($x^{{ld}}_{v,t} = 1$). Within \eqref{eq26}, the summation is over a set of trucks $\Omega_{\left(pr_{n^{depot}},n^{port}\right)}$ eligible to deliver product $pr_{n^{depot}}$ from $n^{port}$ to depot $n^{depot}$\footnote{Here $n^{depot}$ must be equal to $n^{depot}_v$ by assumption 1 stated in the beginning of the Section.} toward satisfaction of demand $D_{n^{depot}}$ if truck $v$ is loaded ($x^{{ld}}_{v,t} = 1$). 


\noindent \textbf{Charging Station Capacity Constraints.} To avoid the situation that more trucks are charging at the same time $p$ at node $n^{chrg}$ than the number of chargers $C_{n^{chrg}}$, the following ``charging station capacity'' constraint is introduced: 
\begin{flalign}
& \sum_{v \in \mathcal{V},t \in \mathcal{T}} 
x^{chrg}_{v,n^{chrg},p,t} \leq C_{n^{chrg}}, \forall (n^{chrg} \in \mathcal{N}^{chrg}, p \in \mathcal{P}). \label{eq27}  
\end{flalign}
In \eqref{eq27}, the summation is over trips and trucks since a given point in time $p$ may correspond to different trips for different trucks. 

\noindent \textbf{Product \textcolor{blue}{Unloading} Constraints and Tardiness.} 
The \textcolor{blue}{unloading start} times $u_{v,t,pr_n}$ that are captured in \eqref{eq16}, are used to determine the latest \textcolor{blue}{unloading} time and tardiness. The latest \textcolor{blue}{unloading} time is determined as:  
\begin{flalign}
& \overline{\textcolor{blue}{u}}_{pr} \geq \textcolor{blue}{u}_{v,t,pr}, \forall (v \in \mathcal{V}, t \in \mathcal{T}, pr \in \mathcal{PR}).  \label{eq28}
\end{flalign}
The tardiness is then defined as: 
\begin{flalign}
& \overline{\textcolor{blue}{u}}_{pr} - due_{pr} \leq tard_{pr}, tard_{pr} \geq 0, \forall (pr \in \mathcal{PR}).  \label{eq29}
\end{flalign}
Tardiness is only positive if the unloading start time is later than the due time, otherwise, the tardiness should be zero. Both situations are materialized through optimization, which is presented ahead. 



\subsection{Objective Function.} The objective function is to minimize the total labor, charging, and tardiness costs as follows: 
\begin{flalign}
& \min \left\{\sum_{v \in \mathcal{V}} O_v(d,\overline{a},x) + \sum_{pr \in \mathcal{PR}} O_{pr}(tard)\right\} =  \min \Bigg\{\sum_{v \in \mathcal{V}} C^{lbr} \cdot \left(\overline{a}_{v,n_{v}^{depot}} - d_{v,n_{v}^{depot},1}\right)  + \label{eq30} \\ & 
\sum_{v \in \mathcal{V}, n^{chrg} \in \mathcal{N}^{chrg}, t \in \mathcal{T}, p \in \mathcal{P}} C^{chrg}_{p,n^{chrg}} \cdot x^{ch}_{v,n^{chrg},p,t} +  \sum_{pr \in \mathcal{PR}} C^{tard}_{pr} \cdot tard_{pr}\Bigg\}. \nonumber 
\end{flalign}
Here $\overline{a}_{v,n_{v}^{depot}}$ is introduced to capture the latest arrival time of vehicle $v$ at the depot $n_{v}^{depot}$ as:
\begin{flalign}
& \overline{a}_{v,n_{v}^{depot}} \geq a_{v,n_{v}^{depot},t}, \forall (t \in \mathcal{T}). \label{eq31}
\end{flalign}
Accordingly, the term $\left(\overline{a}_{v,n_{v}^{depot}} - d_{v,n_{v}^{depot},1}\right)$ is the total time a driver spends on the road (difference between the completion of the last trip taken and the departure during the first trip). 
The total labor and charging costs are collectively represented as an additive form $\sum_{v \in \mathcal{V}} O_v(d,\overline{a},x)$ in terms of trucks $v$. 
The drastic reduction of complexity through the formulation of much smaller and much less complex truck subproblems will be exploited in Section \ref{SolutionMethodology}. 

To avoid the unnecessary increase in complexity, the maximum number of trips as well as the number of time periods need to be appropriately selected as discussed next. 

\subsection{Selection of the Maximum Number of Trips and Time Periods} \label{sec25}

The number of $T$ is generated based on demand and supply. For instance, consider a scenario where the demand at a port is 18, the demand at a warehouse is 12, and there are 10 trucks. To avoid the unnecessary increase in complexity, the number of round trips is set to 2, accordingly, the number of individual trips is 4. One possible solution is that 10 trucks pick up the cargo demanded during their first trip, and 8 trucks need a second trip to pick up the remaining 8 units of cargo. 
Therefore, we use a general formula: $2 \cdot \lceil\frac{\max\{D_{n^{depot}},pr_{n^{depot}},D_{n^{port},pr_{n^{port}}}\}}{V}\rceil$ to determine the maximum number of trips. The quantity under the ``ceiling'' operator is the number of round trips, and the factor of 2 arises to compute the total number of trips.

The determination of the number of time periods follows two main guidelines: to avoid too few time periods, which may result in infeasibility, and too many time periods, which may exacerbate combinatorial complexity. We heuristically estimate the longest path without loops and driving back from a port to a depot. The duration of this path is then multiplied by the largest number of trips defined above.





\section{Solution Methodology} \label{SolutionMethodology}

This section presents a novel use of the Surrogate ``Level-Based" Lagrangian Relaxation (SLBLR) approach
\cite{Bragin22}. This method, enhanced by penalties, 
is employed to optimize the nonconvex dual function as well as to recover primal feasibility, aiming to solve the JRC problem formulated in the previous section \ref{problemformulation}. Our approach was selected to reduce combinatorial complexity and to efficiently coordinate truck subproblems while facilitating faster convergence by employing ``Level-Based'' functionality. 

The algorithm's key strengths lie in its ability to 1) ``reverse'' combinatorial complexity - a desirable feature for complex optimization problems, and 2) employ a ``surrogate'' component that avoids the need for optimal subproblem solutions at each iteration for truck subproblems. This surrogate feature significantly decreases computational effort and alleviates multiplier zigzagging, offering an advantage over conventional LR methods.
Moreover, traditional LR often suffers from coordination challenges - the requirement for non-summable step sizes to ensure convergence, which impedes \textit{iteration-wise} progress. In contrast, our method adopts a decision-based approach to determine level values for stepsizing, thereby efficiently exploiting a linear convergence rate in the dual space without necessitating the non-summability of step sizes. This efficiency is facilitated by an auxiliary problem that promptly identifies and corrects multiplier divergence, leading to accelerated level adjustments.

Empirical testing against simpler problems or those with smaller gaps, as cited in Ref \cite{Bragin22}, demonstrated significant improvement over other dual and primal methods, and is thus a promising approach for the problem formulated in Section \ref{problemformulation}.

In our approach, we relax the coupling constraints that link different trucks, including demand \eqref{eq25}-\eqref{eq26} and charging station capacity constraints \eqref{eq27}, and penalize their violations. By doing so, we will be able to formulate truck subproblems, thereby improving computational efficiency. This type of relaxation is coupled with penalties for constraint violations that promote the accelerated reduction of constraint violations.

The optimal coordination of trucks is tantamount to the optimization of the nonconvex dual function, which, in a general form can be written as: 
\begin{flalign}
& \max_{\Lambda} \{q_\rho(\Lambda): \Lambda \in \mathbb{R}^{|\mathcal{N}^p|\cdot|\mathcal{PR}|} \times \mathbb{R}^{|\mathcal{N}^d|\cdot|\mathcal{PR}|} \times \mathbb{R}^{|\mathcal{N}^c|\cdot|\mathcal{P}|}\},  \label{eq32}
\end{flalign}
with
\vspace{-2mm}
\begin{flalign}
& q_\rho(\Lambda) = \min_{x,a,d,u,s} \Bigg\{L_\rho(x,a,d,u,s,\Lambda)\Bigg\},  \label{eq33}
\end{flalign}
where $L_\rho(x,a,d,u,s,\Lambda) \equiv \sum_{v \in \mathcal{V}} O_v(d,k,x) + O_{pr}(tard)$ + $\Lambda \cdot H(x)$ + $\rho \cdot \|H(x)\|_1$ is the ``absolute-value'' Lagrangian function \cite{Bragin19}. Here $\Lambda$
is the vector of Lagrangian multipliers, variables $x$ above collectively define all the binary decision variables; variables $a,$ $d,$ $u$, and $s$ define the arrival time, departure time, unloading time, and the state of charge decision variables, respectively.


Minimization within \eqref{eq33} is referred to as the ``relaxed problem.'' $H(x)$ is a vector of constraint violations defined as: 
\vspace{-2mm}
\begin{flalign}
& H(x) = \begin{pmatrix}
\sum\limits_{t \in \mathcal{T}^{inbnd} \subset \mathcal{T}, v \in \Omega_{\left(pr_{n^{port}},n_v^{depot}\right)}}   x^{{ld}}_{v,t}  -  \sum\limits_{pr_{n^{port}}  \in  \mathcal{PR}} D_{n^{port},pr_{n^{port}}} \\
\sum\limits_{t \in \mathcal{T}^{otbnd} \subset \mathcal{T}, v \in \Omega_{\left(pr_{n^{depot}},n^{port}\right)}}  x^{{ld}}_{v,t}  -  \sum\limits_{pr_{n^{depot}}  \in  \mathcal{PR}} D_{n^{depot},pr_{n^{depot}}} \\
\sum\limits_{v \in \mathcal{V}, t \in \mathcal{T}} 
x^{chrg}_{v,n^{chrg},p,t} + sl_{n^{chrg},p} - C_{n^{chrg}} \label{eq34}
\end{pmatrix}. 
\end{flalign}
where $sl$ are slack variables used to convert charging station capacity constraints into equality constraints.  

The success of Lagrangian Relaxation methods generally depends on the efficiency of the 
optimization of the non-smooth dual function \eqref{eq32}.

To maximize the dual function (refer to Equation \eqref{eq32}), the SLBLR approach updates the Lagrangian multipliers $\Lambda$ by taking steps $\alpha^k$ along ``surrogate'' subgradient directions $H(\tilde{x}^k)$ (rather than conventional subgradients):
\vspace{-2mm}
\begin{flalign}
& \Lambda^{k+1} = \Lambda^k + \alpha^k \cdot H(\tilde{x}^k).   \label{eq35}
\end{flalign}
The components of $\Lambda$ associated with ``charging capacity'' constraints are projected onto a positive orthant $\left\{\lambda: \lambda \geq 0\right\}$.
To set stepsizes, SLBLR utilizes a decision-driven scheme (see \eqref{eq37} ahead) to determine a series of ``level values'' that are overestimates of $q(\Lambda^*)$.
Following \cite{Bragin22}, the step-sizes are set as by using ``level values'' $\overline{q}_j$ as: 
\vspace{-2mm}
\begin{flalign}
& \alpha^k = \zeta \cdot \frac{1}{V} \cdot \frac{\overline{q}_{j} - L_{\rho}(\tilde{x}^k,\tilde{a}^k,\tilde{d}^k,\tilde{u}^k,\tilde{s}^k,\Lambda^k)}{\left\|H(\tilde{x}^k)\right\|^2_2}, \zeta < 1.  \label{eq36}
\end{flalign}
Here ``tilde'' is used to distinguish optimal solutions to the relaxed problem, from the solutions to the relaxed problem obtained subject to the ``surrogate optimality condition'' (see \eqref{eq40a} ahead). Accordingly, $L_{\rho}(\tilde{x}^k,\tilde{a}^k,\tilde{d}^k,\tilde{u}^k,\tilde{s}^k,\Lambda^k)$ is referred to as a ``surrogate dual value'' which is an optimized (not necessarily optimal) value of the ``absolute-value'' Lagrangian function. 

To operationalize the above scheme, the following is required:  

%
\begin{enumerate}
\item {The values of $\overline{q}_{j}$ are set after detecting divergence of multipliers through the following ``multiplier-divergence-detection'' feasibility problem:
\begin{flalign} 
    \begin{cases}
      \|\Lambda-\Lambda^{k_j+1}\| \leq \|\Lambda-\Lambda^{k_j}\|,\\
      \|\Lambda-\Lambda^{k_j+2}\| \leq 
\|\Lambda-\Lambda^{k_j+1}\|,\\
...\\
\|\Lambda-\Lambda^{k_j+n_j}\| \leq 
\|\Lambda-\Lambda^{k_j+n_j-1}\|. \label{eq37}
    \end{cases}
\end{flalign}
If the above system of equations admits no feasible solution with respect to $\Lambda$ for some $k_j$ and $n_j$, then $\exists \; \kappa \in [k_j,k_j+n_j]$ such that 
\begin{flalign} 
& \overline{q}_{j} = \max_{\kappa \in [k_j,k_j+n_j]} \overline{q}_{\kappa,j} > q_{\rho}(\Lambda^{*}), \label{eq38}
\end{flalign}
where 
\begin{flalign} 
& \overline{q}_{\kappa,j} = \alpha^{\kappa} \cdot V \cdot \|H(\tilde{x}^\kappa)\|^2 + L_{\rho}(\tilde{x}^\kappa,\tilde{a}^\kappa,\tilde{d}^\kappa,\tilde{u}^\kappa,\tilde{s}^\kappa,\Lambda^\kappa). \label{eq39}
\end{flalign}
While $k$ is the iteration number, $j$ is the ``level-value'' update number; the same level value $\overline{q}_{j}$ is used for multiplier iterations $\kappa \in [k_j,k_j+n_j].$ Iterations $k_j$ and $n_j$ are determined \textit{post factum} based on how long it takes to detect multiplier divergence and how often.} 
\item The ``tilde'' is used to denote solutions to the relaxed problem obtained by solving one subproblem at a time rather than carrying out exact minimization within \eqref{eq33}. Subproblems are formulated as follows:   
\begin{flalign}
& \min \{O_v(d,\overline{a},x) + O_{pr}^{k-1}(tard) + \Lambda \cdot H^{k-1}(x) + \rho \cdot \|H^{k-1}(x)\|_1\},  \label{eq40}
\end{flalign}
where 
\vspace{-3mm}
\begin{flalign}
& H^{k-1}(x)  = \begin{pmatrix}
\sum\limits_{t \in \mathcal{T}^{inbnd} \subset \mathcal{T}, v' \in \Omega_{\left(pr_{n^{port}},n_v^{depot}\right)}\setminus \{v\}} \!\! x^{ld,k-1}_{v,t} + x^{ld}_{v,t} \!-\! \sum_{pr_{n^{port}} \in \mathcal{PR}} D_{n_{port},pr_{n^{port}}} \\
\sum\limits_{t \in \mathcal{T}^{otbnd} \subset \mathcal{T}, v' \in \Omega_{\left(pr_{n^{depot}},n^{port}\right)}\setminus \{v\}} \!\! x^{ld,k-1}_{v,t} + x^{ld}_{v,t} \!- \!\sum_{pr_{n^{depot}} \in \mathcal{PR}} D_{n_{depot},pr_{n^{depot}}} \\
\sum\limits_{t \in \mathcal{T}, v' \in \mathcal{V}\setminus \{v\}} 
x^{chrg,k-1}_{v',n^{chrg},p,t} + x^{chrg}_{v,n^{chrg},p,t} + sl_{n^{chrg},p}- C_{n^{chrg}} \label{eq41}
\end{pmatrix} 
\end{flalign}
is a vector of constraint violations and 
\vspace{-1.5mm}
\begin{flalign}
& H(x^k)  = \begin{pmatrix}
\sum\limits_{t \in \mathcal{T}^{inbnd} \subset \mathcal{T}, v' \in \Omega_{\left(pr_{n^{port}},n_v^{depot}\right)}\setminus \{v\}} \!\! x^{ld,k-1}_{v,t} + x^{ld,k}_{v,t} \!-\! \sum_{pr_{n^{port}} \in \mathcal{PR}} D_{n_{port},pr_{n^{port}}} \\
\sum\limits_{t \in \mathcal{T}^{otbnd} \subset \mathcal{T}, v' \in \Omega_{\left(pr_{n^{depot}},n^{port}\right)}\setminus \{v\}} \!\! x^{ld,k-1}_{v,t} + x^{ld,k}_{v,t}\! - \!\sum_{pr_{n^{depot}} \in \mathcal{PR}} D_{n_{depot},pr_{n^{depot}}} \\
\sum\limits_{t \in \mathcal{T}, v' \in \mathcal{V}\setminus \{v\}} 
x^{chrg,k-1}_{v',n^{chrg},p,t} + x^{chrg,k}_{v,n^{chrg},p,t} + sl^k_{n^{chrg},p}- C_{n^{chrg}} \label{eq41}
\end{pmatrix} 
\end{flalign}
is a vector of surrogate subgradient multiplier-updating directions computed after solving subproblem \eqref{eq40}.
Moreover, $O_{pr}^{k-1}(tard)$, which is defined through tardiness $tard_{pr}$, is affected by the unloading constraints as follows: 
\begin{flalign}
& \overline{u}_{pr} \geq u_{v,t,pr}, \forall (t \in \mathcal{T}, pr \in \mathcal{PR}), \nonumber \\
& \overline{u}_{pr} \geq u^{k-1}_{v',t,pr}, \forall (v' \in \mathcal{V}\setminus\{v\}, t \in \mathcal{T}, pr \in \mathcal{PR}). \label{eq42}
\end{flalign}
The difference between \eqref{eq28} and \eqref{eq42} is that within \eqref{eq42}, decision variables associated with electric trucks other than $v$ are fixed at the latest values available up to the previous iteration $k-1$.  
\end{enumerate}

The minimization within \eqref{eq40} involves piecewise linear penalties ($l_1$ norms), that efficiently penalize constraint violations and are exactly linearizable through the standard use of \textit{special ordered sets}, thereby enabling the use of MILP solvers. 

\noindent \textbf{Surrogate Optimization.} Even though the decomposition into truck subproblems exponentially reduces computational effort, the effort required to obtain the exact minima of the subproblems \eqref{eq40} can be significant. To further reduce the effort, optimization of \eqref{eq40} is subject to the following ``surrogate optimality condition,'' which is an extension of that within \cite[eq. (11)]{Bragin22}:
\begin{flalign}
& O_v(\tilde{d}^k,\tilde{\overline{a}}^k,\tilde{x}^k)  + O_{pr}^{k-1}(\tilde{tard}^k) + \Lambda^k \cdot H^{k-1}(\tilde{x}^k) + \rho^k \cdot \|H^{k-1}(\tilde{x}^k)\|_1 \nonumber < \\ & O_v(\tilde{d}^{k-1},\tilde{\overline{a}}^{k-1},\tilde{x}^{k-1}) + O_{pr}^{k-1}(\tilde{tard}^{k-1}) + \Lambda^k \cdot H^{k-1}(\tilde{x}^{k-1}) + \rho^k \cdot \|H^{k-1}(\tilde{x}^{k-1})\|_1.  \label{eq40a}
\end{flalign}
Essentially, we only seek a solution that improves upon the subproblem incumbent solution, as found up until the previous interaction $k-1$, but taking into account the updated values of multipliers $\Lambda^k$ and penalty coefficients $\rho^k$. This approach significantly further reduces the associated computational effort. 

\noindent \textbf{Algorithm.} The entire algorithm is summarized as follows:

\textbf{Input} $\Lambda^0,$ $\rho^0,$ 
$\beta > 1$, $\zeta < 1,$ $v = 1$, $k = 1$, $\overline{q}_0 = +\infty,$ $q^{max} = -\infty$

1. \textbf{While} stopping criteria are not satisfied \textbf{do} 

2. Solve subproblem \eqref{eq40}, 

3. Calculate $H(\tilde{x}^{k}),$

4. Calculate $L_{\rho}(\tilde{x}^k,\tilde{a}^k,\tilde{d}^k,\tilde{u}^k,\tilde{s}^k,\Lambda^k),$

5. Calculate stepsizes per \eqref{eq36}, 

6. Update multipliers per \eqref{eq35}, 


7. \textbf{If} $q^{max} < \alpha^{k} \cdot V \cdot \|H(\tilde{x}^k)\|^2 + L_{\rho}(\tilde{x}^k,\tilde{a}^k,\tilde{d}^k,\tilde{u}^k,\tilde{s}^k,\Lambda^k)$ \\ \textbf{then} $q^{max} = \alpha^{k} \cdot V\cdot \|H(\tilde{x}^k)\|^2 + L_{\rho}(\tilde{x}^k,\tilde{a}^k,\tilde{d}^k,\tilde{u}^k,\tilde{s}^k,\Lambda^k)$

\quad \textbf{EndIf}

8. $v \leftarrow v + 1$

9. $k \leftarrow k + 1$

10. \textbf{If} $v = V$ \textbf{then} $v = 1$

\quad \; \textbf{EndIf}

11. \textbf{If} \eqref{eq37} is infeasible \textbf{then} $\overline{q}_j = q^{max}, q^{max} = - \infty$, $j \leftarrow j + 1$ 

\quad \; \textbf{EndIf}

12. \textbf{If} $H(\tilde{x}^k) = 0$ \textbf{then} record feasible cost \textbf{and} $\rho^k \leftarrow \rho^{k-1}/\beta, \beta > 1.$

\quad \; \textbf{EndWhile}

\noindent \textbf{Brief Discussion of the Algorithm.} In 1, the following criteria can be used: number of iterations, CPU time, or a duality gap. To obtain a duality gap, a feasible cost and a dual value are needed. The feasible cost is obtained per 12, whereas to obtain a dual value, additional effort is needed: the penalty coefficient $\rho$ needs to be set to 0, and all electric-truck subproblems need to be solved to optimality.

\section{Numerical Testing} \label{NumericalTesting}

The solution methodology is implemented in CPLEX 22.1.0.0 with default settings on an Intel(R) Core(TM) i9-9900X 3.50 GHz server with 64 GB RAM. In Example 1, a small instance with 5 trucks, 1 port, and 1 warehouse is considered. The purpose is to compare the performance of the new method with the performance of CPLEX \textcolor{blue}{as well as to test the stability of the method}. In Example 2, a medium-sized instance with 15 trucks, 1 port, and 1 warehouse is considered. The purpose is to demonstrate that JRC leads to much reduced operational costs as compared to scheduling that considers shortest paths. In Example 3, a large-scale real-world problem in the Greater Los Angeles area with 50 trucks, 1 port, and 3 depots is considered to demonstrate the scalability of the method. Furthermore, Example 3 examines various case studies with different parameters including the number of charging stations, charging speed, and battery capacity size for electric trucks to demonstrate the economic impacts of these variables. 

\subsection{Example 1: Small Cases with up to 5 Trucks.} 


This example aims to: 1. Show how combinatorial complexity increases with more trucks, trips, and the units of demand; 2. Reveal that an increase in trucks to five renders conventional solvers unable to find feasible solutions; \textcolor{blue}{and 3. Confirm the SLBLR method's stability.} We compare SLBLR to standard Lagrangian Relaxation, which updates multipliers via subgradient directions and adjusts step sizes based on levels \cite{goffin1998convergence}, to ensure a fair assessment.

\begin{figure}[H]
  \centering
    \includegraphics[trim=0 0 0 0,  width=0.57\linewidth, scale=0.5]{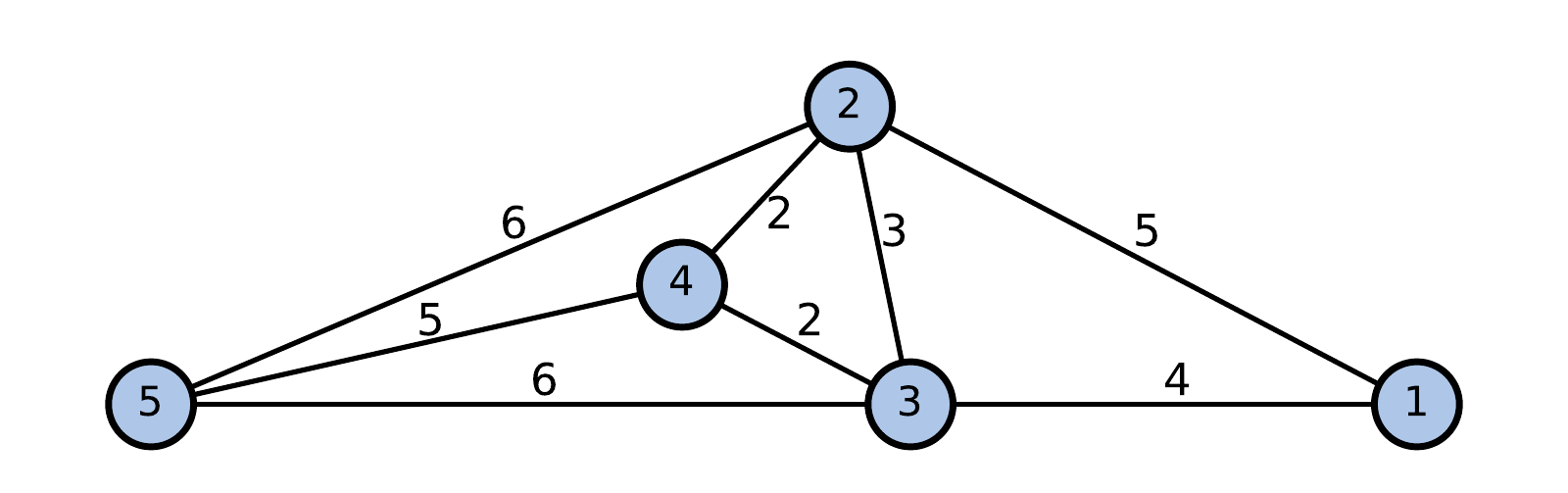}
    \caption{Network topology for Examples 1 and 2. The number of time periods required to travel (in both directions) is shown next to the respective road segments.}
    \label{fig0.5}
\end{figure}

In this example, a transportation network with 5 nodes (shown in Figure \ref{fig0.5}) is considered. Road segments are labeled with travel times, assumed equal in both directions. The charging rate is 17\% per period, with discharge rates at 5\% when loaded and 2.5\% when not. Charging station capacities at each node are ${2, 2, 1, 2, 2}$. Cargo transport due times are 45 steps for import and 20 for export.

The first case contains one truck having to transport one unit of cargo to the port and one unit of cargo to the depot on the way back. As demonstrated in Table \ref{Ex1cases} (first row), this case is solved within a matter of a couple of seconds. As the demand and the number of trips double (second row),
the CPU time is 130.38 seconds - a 58-time increase. As the demand and the number of trucks double (third row), 
the CPU time increases to 70,134.89 seconds, thereby constituting another 537-fold increase over the results presented in the second row. This is a signature behavior of combinatorial NP-hard discrete optimization problems - as the size of the problem increases linearly, the resulting complexity increases superlinearly (e.g., exponentially).

\begin{table}[!h]
\caption{Illustration of the complexity increase with the increase of the demand, the number of trips, and the number of trucks.}
\begin{tabular}{ c c c c c c c c c c c }
\hline 
$v$	& $t$	& Demand 	& Demand 	& Constraints	& Binaries	& Nonzeros	& Time periods	& CPU Time 	& Gap \\ 
& 	&  at $n^{port}$	&  at $n^{depot}$	& 	& 	& 	& 	 & (Sec)	& (\%)\\ 
\hline 
1	& 2	& 1	& 1	& 10,485	& 2,958	& 289,418	& 65	& 2.23	& 0 \\
1	& 4	& 2	& 2	& 18,474	& 5,447	& 478,529	& 65	&  130.38	& 0 \\
2	& 4	& 4	& 4	& 36,060	& 10,680	& 943,043	& 65	&  70,134.89	& 0 \\
5	& 4	& 3	& 8	& 151,791	& 28,312	& 4,782,805	& 65	&  1,000,000	& No sln. found \\
\hline 
\end{tabular} \label{Ex1cases}

\end{table}

Ultimately, consider the case where five trucks are assigned to manage both 3 and 8 units of demand at the port and depot respectively. This logistics scenario requires a two-step transportation process. Initially, three units of cargo need to be transported from Node 1 (the depot) to Node 5 (the port) for the purpose of export. Following the export process, there is a requirement to pick up eight units of cargo from Node 5 and transport them back to Node 1 for import. Despite the significant computational effort, the problem remains unsolved after 1,000,000 seconds of CPU time.



As described in subsection \ref{sec25}, the number of trips for the original Example 1 (which is also reported in the last row of the table above) is estimated to be 4 and the longest path from a depot to a port is estimated to be 15 ($1 \rightarrow 2 \rightarrow 3 \rightarrow 4 \rightarrow 5$). Therefore, the number of periods is 60. An additional 5 units of time is heuristically added as a cushion. For other newly added scenarios, the number of time periods is kept the same while varying other parameters like the number of trips, vehicles, and demand/supply in order to see how these parameters affect the complexity.


Figure \ref{fig3} demonstrates the performance of the SLBLR method. The first feasible solution found by the SLBLR method is obtained within less than 100 seconds. The method intrinsically possesses the ability to systematically improve the solution, since the ``price-based'' and systematical update of the Lagrangian multipliers leads to ``more economical'' solutions. Moreover, subproblem solutions, when close to feasibility, are easily repaired to obtain feasible solutions. As a result, the method is capable of improving upon previously obtained solutions with ease. Ultimately, a high-quality solution with a gap of less than 1\% is obtained within 5 minutes.

\noindent \textbf{Comparison against other methods.} We compare our method against standard package CPLEX and standard Lagrangian Relaxation (LR) with subgradient directions used to update multipliers. 

\begin{enumerate}

\item \textbf{Comparison against CPLEX:} In attempting to solve the same problem instance, CPLEX is unsuccessful in finding a feasible solution even after 1,000,000 seconds. This puts SLBLR ahead by 4 orders of magnitude in terms of speed. When analyzing lower bounds and solution quality, CPLEX could only compute a lower bound of 3.1452 within the time limit, failing to find a feasible solution. This implies that even with the help of cutting and branching, CPLEX could not make a significant impact on improving the lower bound. It is essential to note that while there is a significant difference between the CPLEX lower bound and ours, we attribute this to the distinct methodologies and heuristics each system employs. As a result, these varying approaches can lead to different lower bounds and overall performances.

\begin{figure}[H]
  \centering
    \includegraphics[trim=0 0 0 0,  width=0.5\linewidth, scale=0.5]{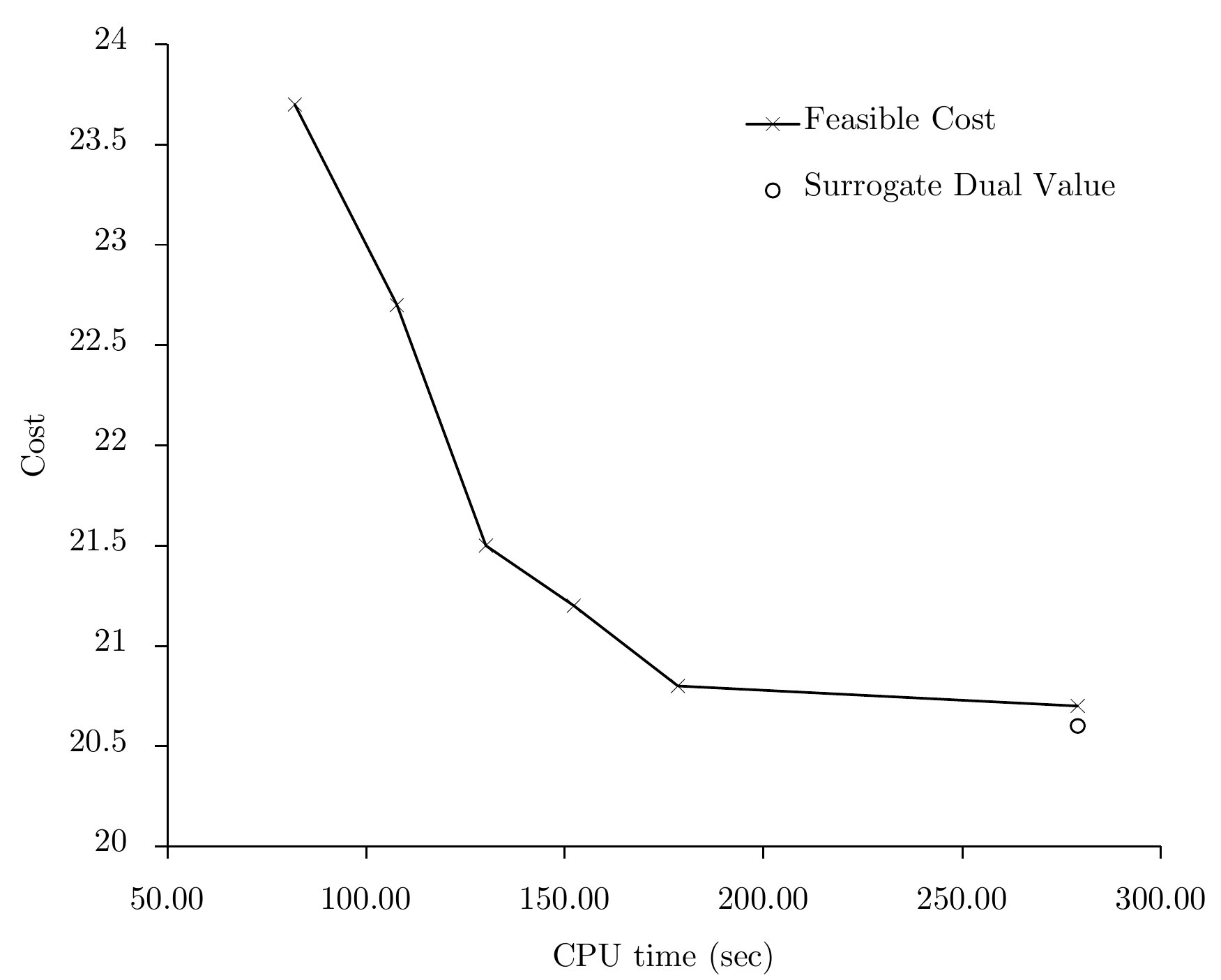}
    \caption{Solutions obtained by the SLBLR method for Example 1}
    \label{fig3}
\end{figure}

\item \textbf{Comparison against LR with Subgradient-Level approach \cite{goffin1998convergence}\footnote{Subgradient-Level method \cite{goffin1998convergence} is chosen because, like our method, it also uses the Polyak formula \cite{Polyak69}, albeit, with a different technique to obtain level values.} to update multipliers:} While LR's strength is the exploitation of the drastic reduction of complexity upon decomposition, standard LR requires that 1. truck subproblems are solved to optimality and 2. all truck subproblems are solved before updating multipliers. In contrast, our method requires subproblem-feasible solutions that satisfy the ``surrogate optimality condition'' which only requires that only one solution that can improve the subproblem incumbent is obtained. Our method requires only 8 seconds to solve a truck subproblem, whereas standard LR requires 566 seconds. After 22,656.31 seconds, LR obtained a feasible solution with a cost of 24 and a lower bound of 18.6 (which is a gap of 22.5\%), whereas our method found a solution with a cost of 20.7 and a lower bound of 20.6 (which is a gap of 0.5\%) after 287 seconds. The results are shown in Figure \ref{SLBLRvsLR}. While the standard Lagrangian Relaxation is more successful than CPLEX, which does not exploit the reduction of complexity upon decomposition, our method is faster than standard LR by roughly 2 orders of magnitude.


\end{enumerate}

\begin{figure}[H]
\centering
\includegraphics[width=0.45\textwidth, angle =0 ]{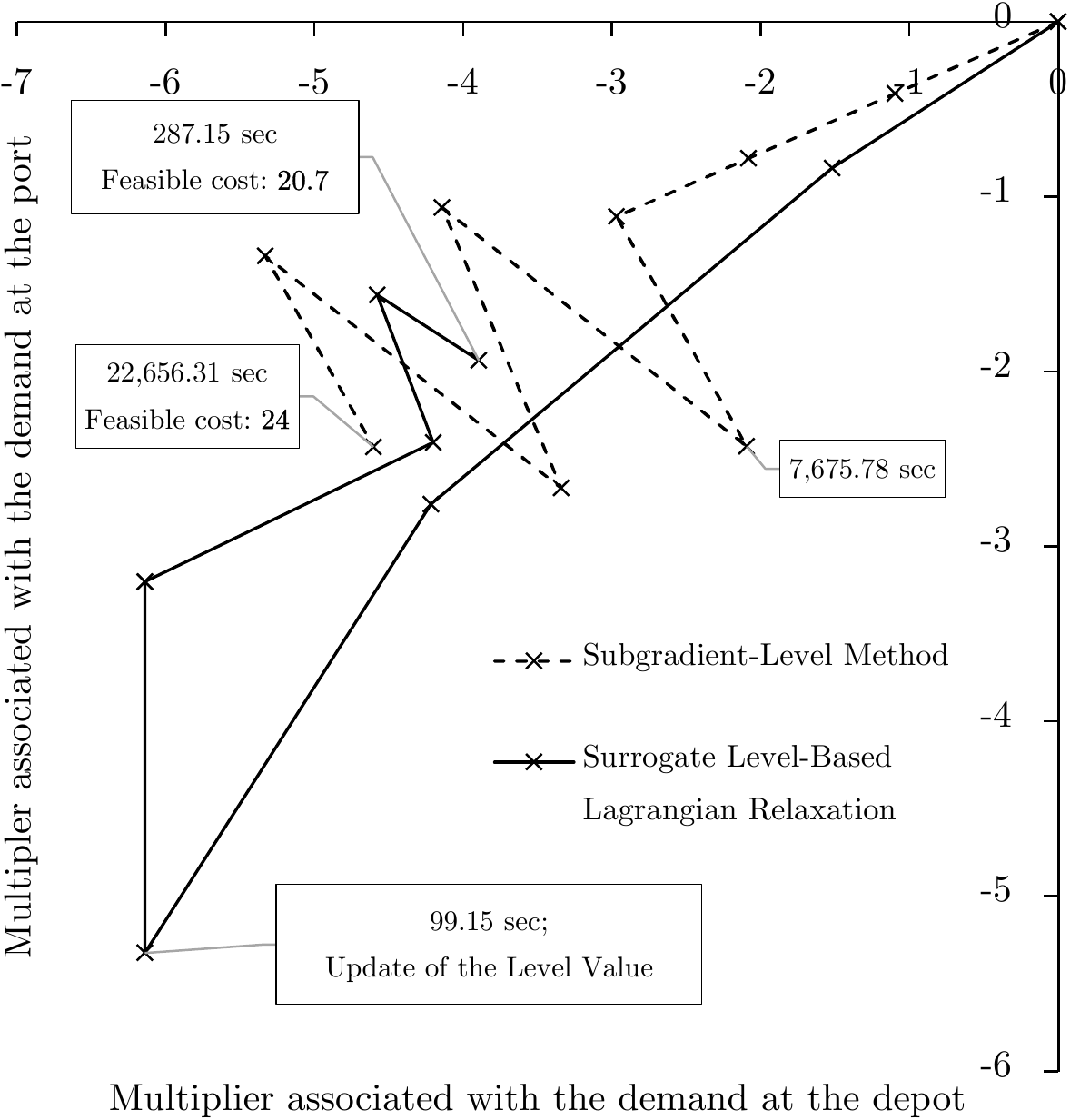}
\caption{Trajectories of multipliers within Surrogate ``Level-Based'' Lagrangian and standard Lagrangian Relaxation with Subgradient-Level method for updating multipliers.}
\label{SLBLRvsLR}
\end{figure}


Figure \ref{SLBLRvsLR} demonstrates trajectories of multipliers generated by standard LR and our method. Within standard LR, multipliers zigzag whereas directions obtained by our method are much smoother. More importantly, our method requires much less time per multipliers update as discussed above.

\noindent \textbf{Stability of SLBLR.} \textcolor{blue}{The stability with respect the coordination-related hyperparameter $\zeta$ was previously established in the foundational study detailed in \cite{Bragin22}. Consistent with this, we have maintained $\zeta$ at a constant value of $1/2$ throughout all Examples. However, the original investigation \cite{Bragin22} did not explore the stability of the primal-solution-recovery parameters, namely $\beta$ and $\rho_0$. In this context, we find it imperative to extend the stability analysis to these parameters and, hence, we present our findings herein.}

\textcolor{blue}{For $\rho_0$ and $\beta$, the baseline values used to obtain results above are 5 and 1.05 respectively. The results of the experiments when varying these parameters are provided in the subsequent tables:}

\begin{table}[h!] 
\caption{CPU times for varying values of  $\beta$.}
\centering
\begin{tabular}{|c|c|c|c|}
\hline
\textbf{$\rho_0$} & \textbf{$\beta$} & \textbf{CPU Time (sec)} \\
\hline
5 & 1.1 & 269 \\
5 & 1.05 & 287 \\
5 & 1.025 & 420 \\
\hline
\end{tabular}
\label{StabWRTBeta}
\end{table}

\begin{table}[h!]
\caption{CPU times for varying values of  $\rho_0$.}
\centering
\begin{tabular}{|c|c|c|}
\hline
\textbf{$\rho_0$} & \textbf{$\beta$} & \textbf{CPU Time (sec)} \\
\hline
10 & 1.05 & 313 \\
5 & 1.05 & 287 \\
2.5 & 1.05 & 314 \\
\hline
\end{tabular}
\label{StabWRTRho}
\end{table}

\textcolor{blue}{According to Tables \ref{StabWRTBeta} and \ref{StabWRTRho}, results are fairly stable. 
The computational performance was evaluated based on the time required to achieve a solution with a cost of 20.7—the near-optimal result achieved in our experiments as explained above. }


\textcolor{blue}{Stability concerning initialization procedures was addressed in the seminal SLBLR paper \cite{Bragin22}. Recognizing that initialization stability may vary with the nature of the problem, our approach included a stability test via randomized multiplier initialization. Multipliers were uniformly distributed within the interval $[-50, 50]$. The CPU times for this process are summarized in Table \ref{randommultrobust}.}

\begin{table}[H]
\caption{CPU times for randomized initialization of multipliers.}
\centering
\begin{tabular}{|c|c|}
\hline
\textbf{Test Instance} & \textbf{CPU Time (sec)} \\
\hline
1 & 446 \\
2 & 472 \\
3 & 399 \\
4 & 324 \\
5 & 308 \\
6 & 362 \\
7 & 524 \\
8 & 302 \\
9 & 331 \\
10 & 251 \\
\hline
Average & 372 \\
Standard Deviation & 86  \\
\hline
\end{tabular}
\label{randommultrobust}
\end{table}
\textcolor{blue}{As compared to a CPU time obtained to solve the instance of Example 1 whereby the multipliers are all initialized at 0, which is 287 seconds, the above average time is 29.72\% since some of the random multipliers may fall far from the actual optimum. Overall, given the random nature of this testing, the increase in the CPU time is rather modest.} 

\begin{figure}[h!]
\centering
\includegraphics[width=1\textwidth, angle =0 ]{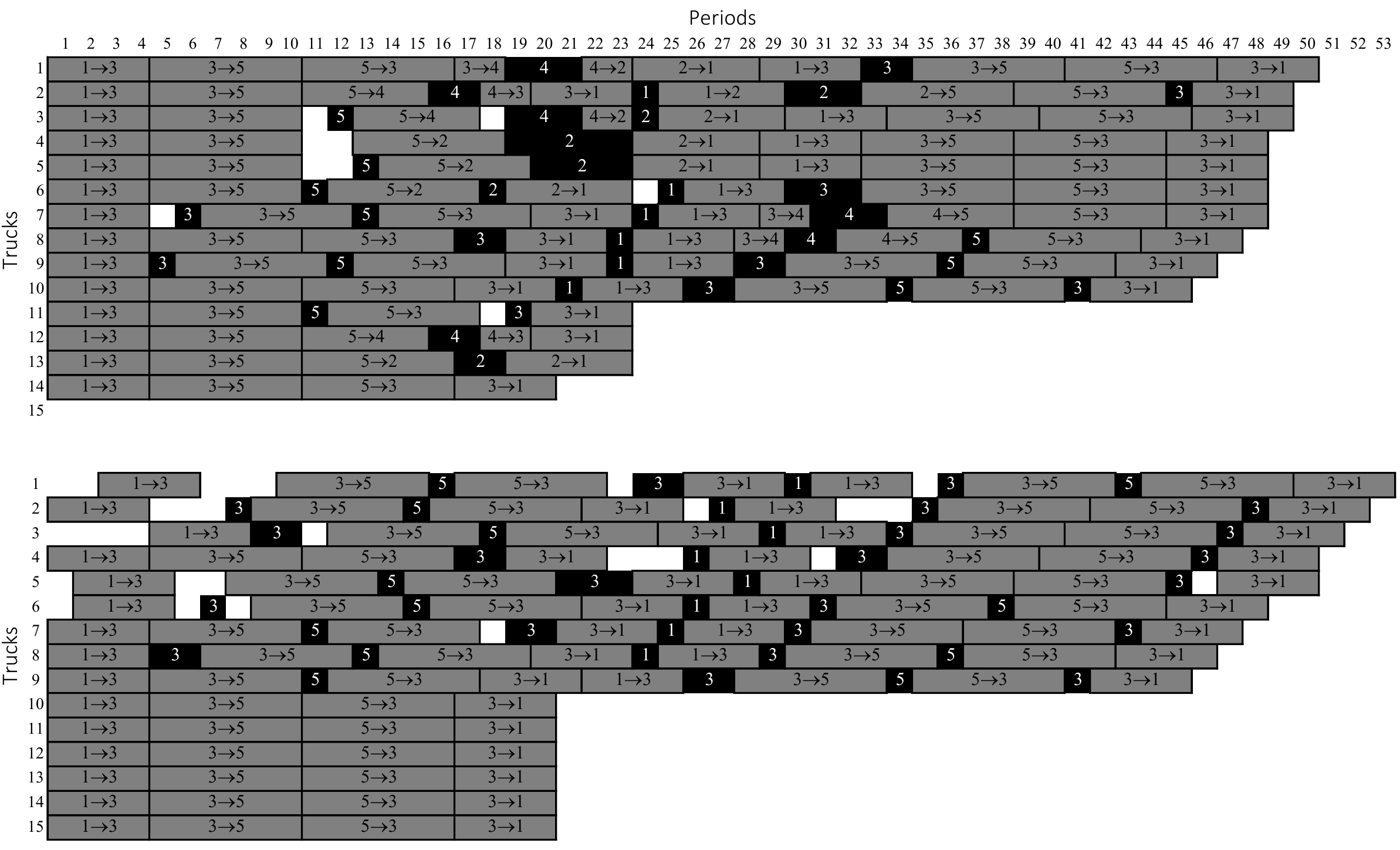}
\caption{A side-by-side comparison of schedules obtained by using the novel modeling versus the ``shortest path'' approach for Example 2.}
\label{Ex2}

\end{figure}

\subsection{Example 2: Medium-Scale Case with 15 Trucks.}

In this example, the same road network is shown in Figure \ref{fig0.5} as well as charging station capacities and due times. The goal of a fleet of 15 trucks is to deliver (to export) 8 units of cargo from node 1 (Depot) to node 5 (Port) and to pick up (to import) 24 units of cargo from node 5 and deliver them to node 1. In order to provide insights into the advantages of joint electric truck routing and charging, results obtained by using the novel formulation are compared with the results obtained while considering shortest paths by disregarding nodes 2 and 4, since visiting these nodes increases the distance and the time traveled. For both cases, the CPU time limit was set to 10 minutes; for the joint scheduling the cost is 68.2 and for the ``shortest path'' scheduling the cost is 73.4. The scheduling results are demonstrated in Figure \ref{Ex2}. Within JRC, since the trucks can take detours through nodes 2 and 4, they save time needed to wait for the availability of the charging station. Ultimately, the overall time (and tardiness) decreases.

For example, trucks 1 and 3 (Figure \ref{Ex2} (top)) take a detour through node 4 to charge, and trucks 4 and 5 take a detour through node 2 since chargers at node 3 are occupied by trucks 8 and 11. In contrast, truck 1 (Figure \ref{Ex2} (bottom)) needs to wait 6 time periods during the first round trip to charge at node 3 because the charger is successively occupied by trucks 4, 7, and 5. Likewise, truck 2 needs to wait three time periods during both the first and the second round trips because the charger at node 3 is occupied by trucks 6 and 8 (during the first trip) and by trucks 3 and 4 (during the second trip). Overall, not only does joint routing and charging lead to a decrease in the cost but also one truck is spared (truck 15). We also discovered that the total time trucks wait for the available charger is reduced by 69\%, and the overall operational cost is reduced by 6.8\%.

\subsection{Example 3: Large-Scale Cases with 50 Trucks in the Greater Los Angeles Area.} 

\begin{figure}[H]
\centering
\includegraphics[width=1\textwidth]{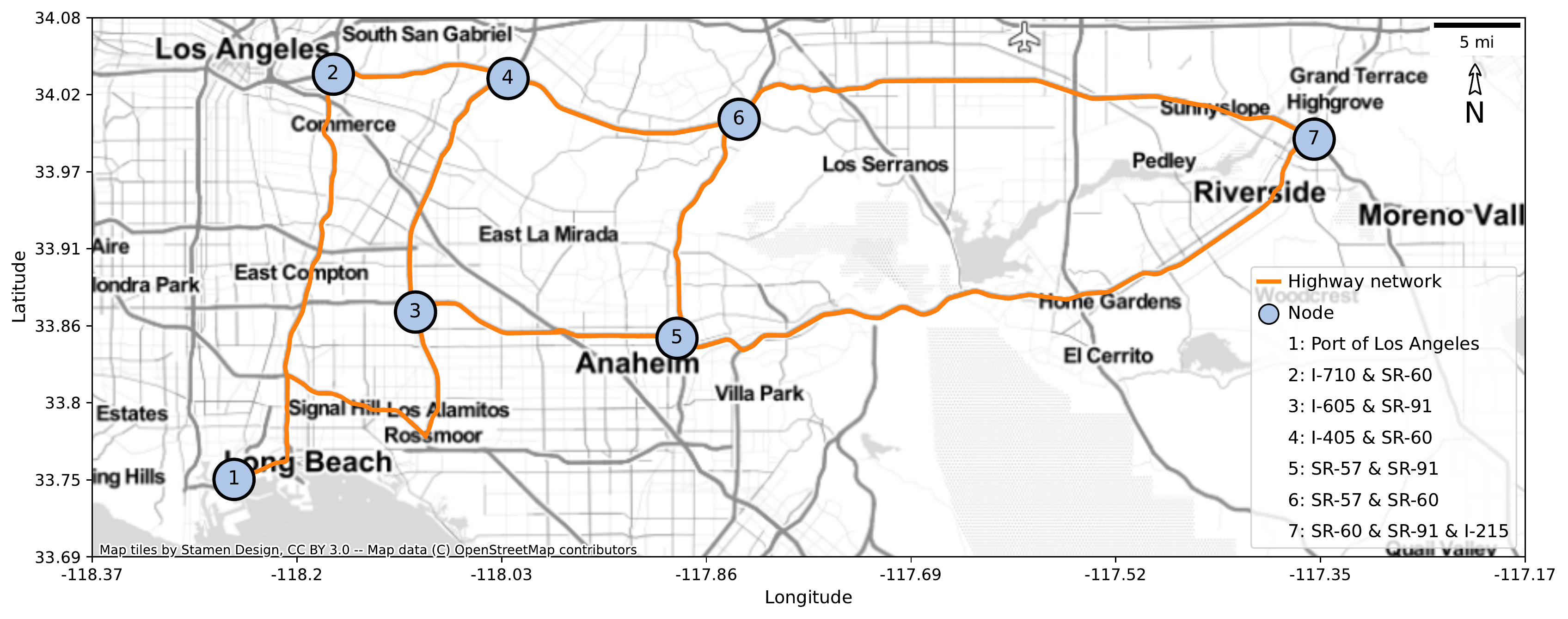}
\caption{Road network, port, warehouses, and charging stations for Example 3.}
\label{fig:overview_old}
\end{figure}

In this example, a realistic road network topology within the Greater Los Angeles area with one port located at Long Beach and three warehouses located at node 4 (intersection of I-405 and SR 60), node 6 (intersection of SRs 57 and 60), and node 7 (intersections of SRs 60 and 91 with I-215) is adopted as shown in Figure \ref{fig:overview_old}.

A fleet of 50 trucks operates from warehouses at nodes 4, 6, and 7, containing 17, 13, and 20 trucks, respectively. The trucks are equipped with battery sizes of either 250 kWh \cite{weiss2020energy,sato2022experimental} or 600 kWh (estimated from \cite{tesla2022semi}), featuring standard and long-range versions. The charging power of the base case is assumed to be 350 kW \cite{deb2021review}. We follow the regression model in \cite{weiss2020energy} to determine energy consumption rates of fully-loaded and empty trucks. Assuming that the weight of a fully loaded and an empty truck is 80,000 lb and 22,000 lb \cite{sato2022experimental}, respectively, we can obtain energy consumption rates of 2.267 kWh/mile for a fully loaded truck and 1.617 kWh/mile for an empty truck. We will study the impact of higher charging powers considering the industry trends \cite{wang2021grid}. The import requirements are 39, 35, and 33, and the export requirements are 32, 20, and 25, respectively. A number of testing cases are considered in this example. 

\noindent \textit{Base Case}: In this case, 30 trucks with 600 kWh batteries are located at the City of Industry and Riverside, and 20 trucks with 250 kWh batteries are located at El Monte. There are 3 chargers available at each node with a maximum charging power of 350 kW. The electric trucks associated with depots 6 and 7 have a longer range with a battery capacity of 600 kWh, while the electric trucks associated with node 4 have a smaller battery capacity of 250 kWh. 

\noindent \textit{Cases 1.1-1.7}: In this series of cases, we test the impacts of the number of chargers on the total operation cost. Accordingly, the number of chargers at nodes 1-7 is reduced from 3 to 1. 

\noindent \textit{Case 2}: In this case study, we evaluate the impacts of the size of the electric truck battery on the total operation cost. It is assumed that all trucks are equipped with a battery capacity of 600 kWh.

\noindent \textit{Case 3}: In this case study, we quantify the impacts of maximum charging power on the total operation cost. The maximum charging power of all chargers is increased from 350 kW to 700 kW.

The optimization stopping criterion is 1800 seconds. The results for all the case studies are reported in Table \ref{table0.5a}.

When the number of chargers is reduced from 3 to 1, the overall operation cost increases the most in Cases 1.2 (8.6\%), 1.3 (7.8\%), 1.1 (7.3\%), and 1.4 (6.7\%) compared to the Base case. This means that the ordering of the best locations for electric truck charging stations is node 2 (intersection of I-710 and SR 60), node 3 (intersection of I-605 and SR 91), node 1 (Port of Long Beach), and node 4 (intersection of I-405 and SR 60). When we compare Case 2 with the Base Case, the increase in battery capacity for the 20 trucks at El Monte from 250 kWh to 600 kWh leads to a reduction in the total operation cost by 27.9\%. Additionally, this enhancement reduces the requirement for the number of electric trucks from 49 to 43. Finally, comparing Case 3 and the Base case, we discover that increasing the maximum charging power of all chargers from 350 kW to 700 kW decreases the operation cost by 9.2\%.
\begin{table}[h!]
\caption{The Feasible Operation Cost and CPU Time of Test Cases 1-3.} 
\centering 
\begin{tabular}{c c c c} 
\hline\hline 
 Case & Feasible  & CPU  & Number of   \\
 & Cost & Time (Sec) & Trucks Used  \\[0.5ex] 
\hline 
Base	&	5527.39	&	894.53	&	49	\\
Case 1.1	&	5934.63	&	1151.99	&	49	\\
Case 1.2	&	6005.08	&	1149.49	&	49	\\
Case 1.3	&	5959.69	&	887.32	&	49	\\
Case 1.4	&	5900.92	&	929.83	&	49	\\
Case 1.5	&	5578.95	&	1150.93	&	49	\\
Case 1.6	&	5683.47	&	1662.93	&	49	\\
Case 1.7	&	5574.17	&	1569.50	&	49	\\
Case 2	&	4020.51	&	1153.15	&	43	\\
Case 3	&	5017.30	&	1327.38	&	48	\\
\hline 
\end{tabular}
\label{table0.5a} 
\end{table}

\begin{figure}[H]

\centering
\includegraphics[width=0.6\textwidth, angle =0 ]{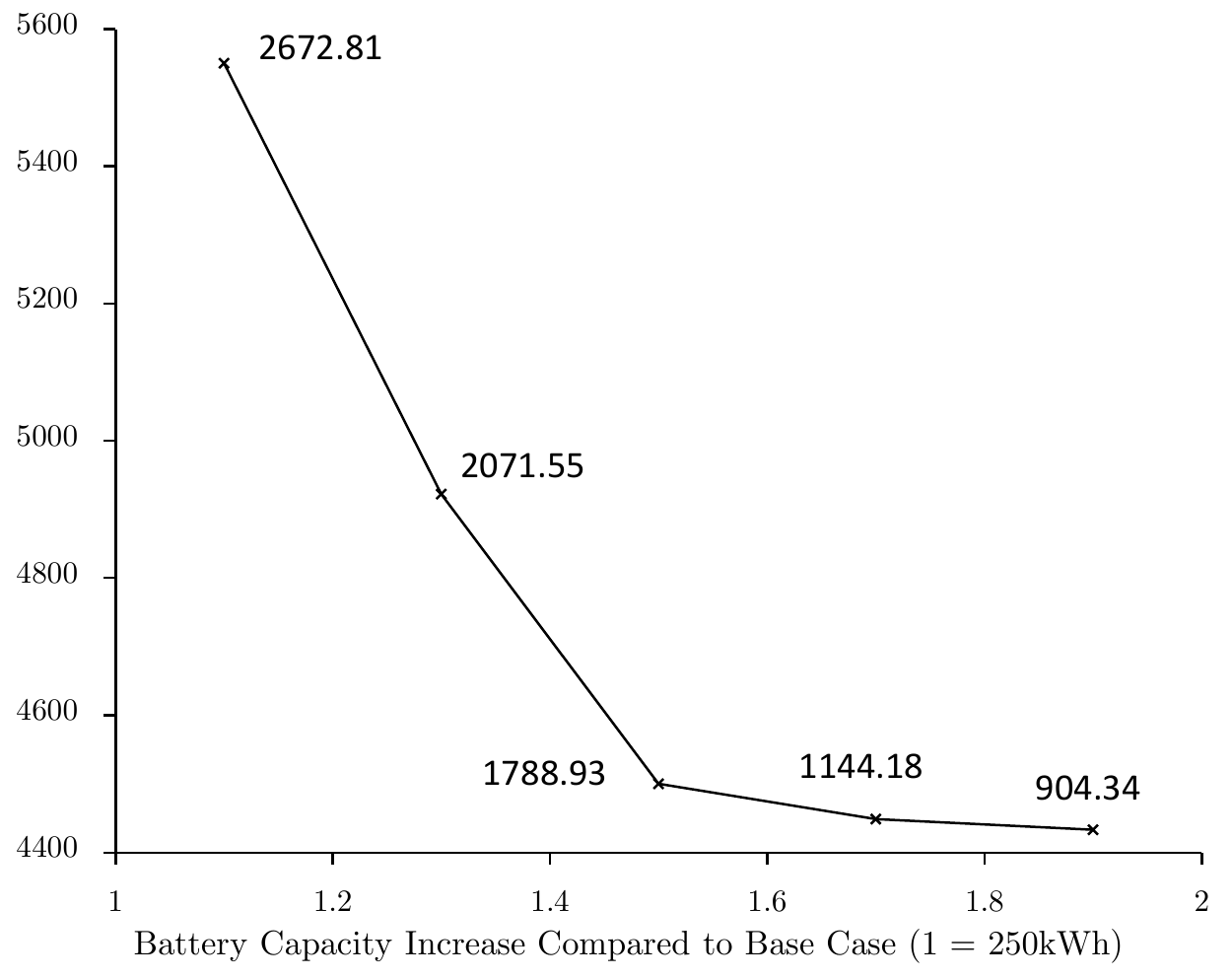}
\caption{Total cost (y-axis) dependence on truck battery capacity. CPU times (in seconds) are also shown next to data points.}
\label{Ex31}

\end{figure}
\noindent \textbf{Analysis of the Impacts of Battery Capacity and Maximum Charging Power on the Cost and CPU Time.} 

In this part of the Example, a network illustrated in Figure \ref{fig:overview_old} is considered, and a fleet of 50 trucks with 250 kWh batteries is assumed. The fleet operates across nodes 4, 6, and 7, housing 10, 20, and 20 trucks respectively. The base case assumes a charging capacity of 350 kW with six chargers per node, import demands of 29, 21, and 15, as well as export demands of 25, 23, and 14, with due times set at 22 and 9 time periods respectively.

The analysis explores the influence of varying battery capacities and charging powers on operational costs and solving times. Fixing the charging power at 110\% (385 kW) of its nominal value, battery capacities of 110\%, 130\%, 150\%, and 170\% (275 kWh, 325 kWh, 375 kWh, and 425 kWh) are considered. Findings, depicted in Figure \ref{Ex31}, indicate negligible cost benefits beyond a 170\% battery capacity increase.

\begin{figure}[H]

\centering
\includegraphics[width=0.60\textwidth, angle =0 ]{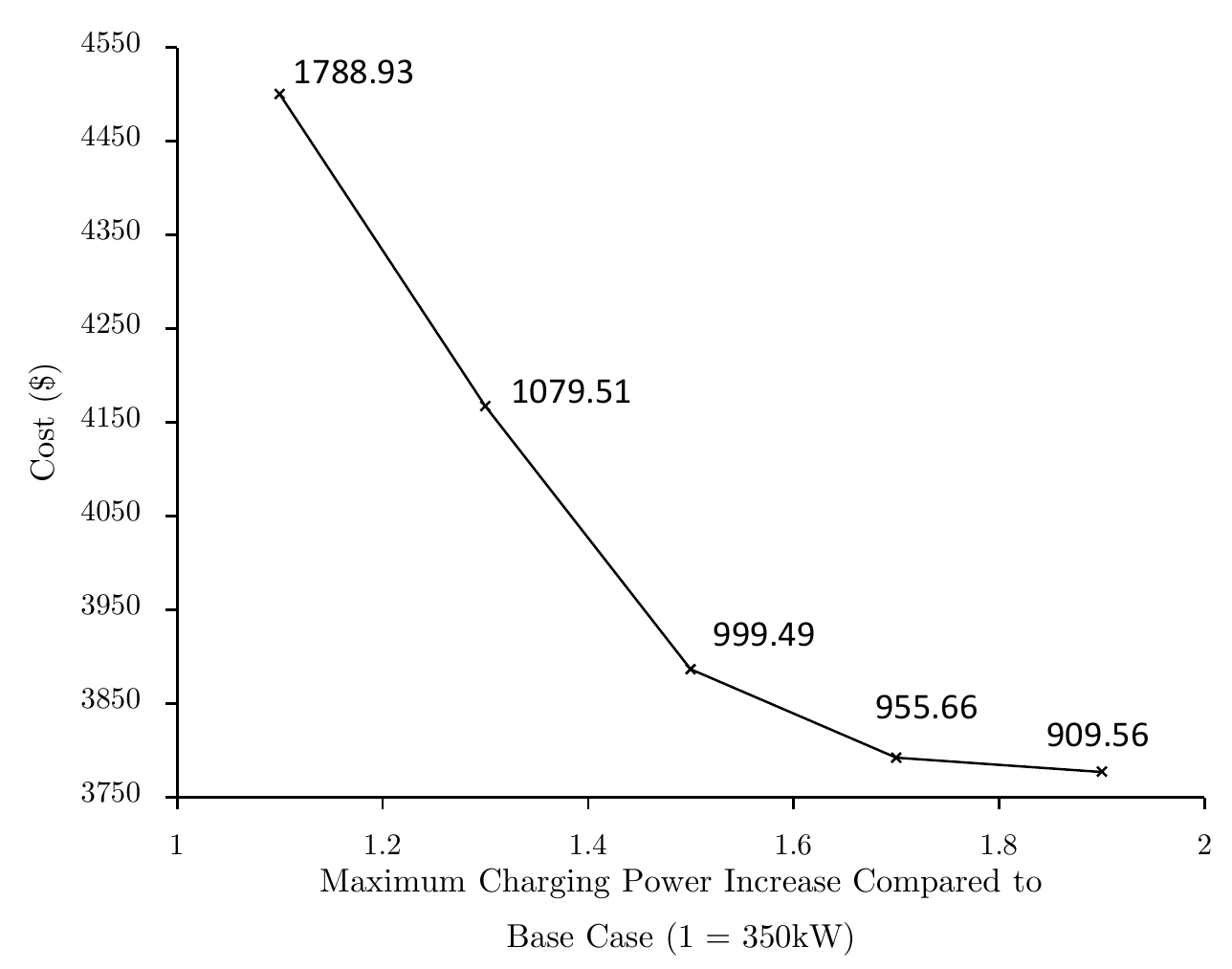}
\caption{Total cost (y-axis) dependence on the maximum charging power of charging stations. CPU times (in seconds) are also shown next to data points.}
\label{Ex32}
\end{figure}

Similarly, with batteries set at 150\% (375 kWh) of nominal capacity, charging powers of 110\%, 130\%, 150\%, and 170\% (385 kW, 455 kW, 525 kW, and 595 kW) are considered. Results (Figure \ref{Ex32}) suggest that increasing charging power above 170\% yields no substantial cost advantage. Notably, enhanced battery capacity and charging power correlate with reduced problem-solving time, as longer-range trucks and higher-powered chargers lead to fewer charging events and thus lower the computational complexity.

\noindent \textbf{Discussion on Solution Quality.} 
Table \ref{table-gap} demonstrates the computation time required to obtain lower bounds to reach the gap of at least 1\% for the feasible costs for five different cases that are demonstrated in Figure \ref{Ex32}.

\begin{table}[!h]
\caption{\textcolor{blue}{Computation Time Required to Obtain Lower Bounds}} 
\centering 
\begin{tabular}{ c c c }
\hline 
Maximum Charging Power Increase & Time (Sec) & Gap (\%) \\
\hline 
1.1	& 111,140	& 1 \\
1.3	& 104,752	& 1\\
1.5	& 101,110	& 1\\
1.7	& 33,774	& 1\\
1.9	& 33,027	& 0.13\\
\hline 
\end{tabular}
\label{table-gap}
\end{table}


While feasible costs are quickly obtained due to the four advantages highlighted in the Introduction (decomposition, surrogate optimization, rapid coordination, and penalization), computing lower bounds for dual values, which provide lower bounds, is more complex. Exact optimality in truck subproblems is necessary for these calculations. The benefits of surrogate optimization for obtaining primal solutions do not extend to lower bound calculations, leading to longer CPU times for solution quality verification. Generally, achieving a satisfactory lower bound quickly is challenging across various problems, with few exceptions. The reduction of CPU time for robust lower bound computation remains a significant challenge and is beyond the scope of this paper. 

\begin{figure}[H]
\centering
\includegraphics[width=1\textwidth]{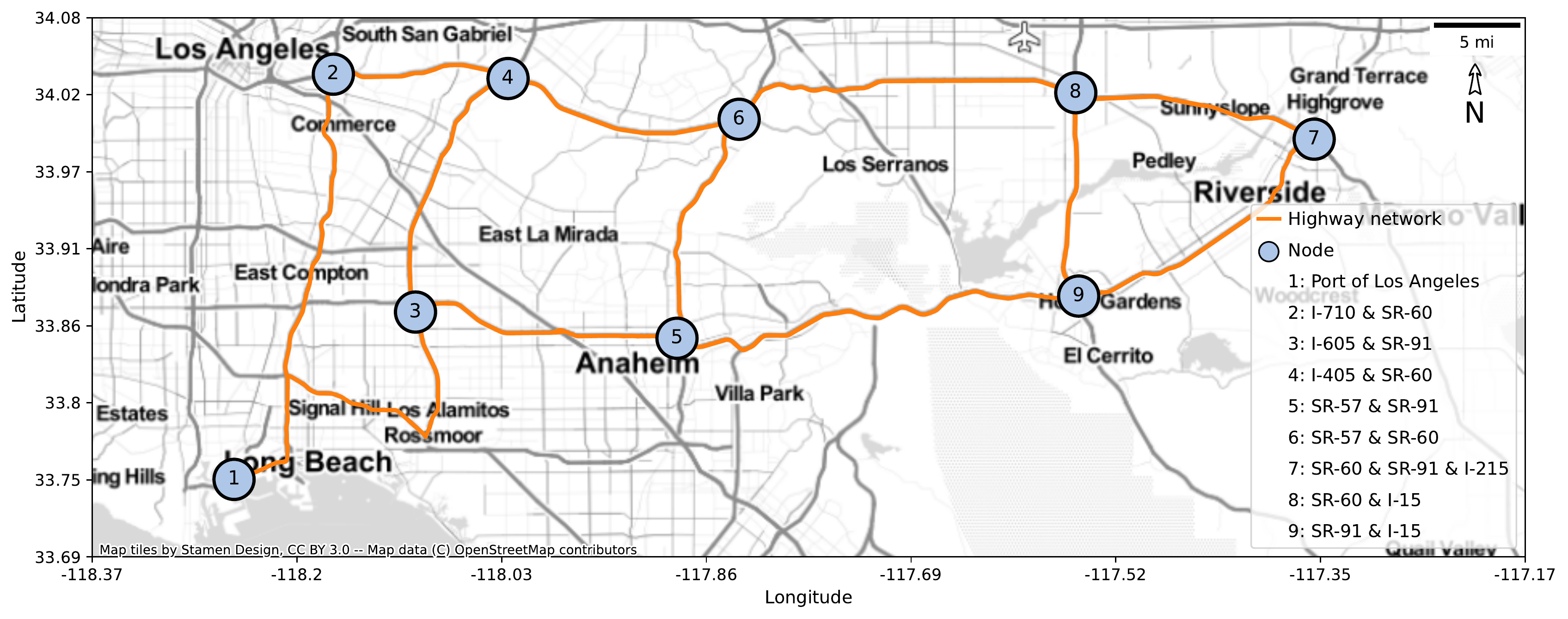}
\caption{An augmented road network, port, warehouses, and charging stations for Example 3.}
\label{fig:overview}
\end{figure}

\noindent \textbf{Scalability with Respect to the Number of Nodes.} The above results demonstrate scalability with respect to the number of trucks. Because of the NP-hardness, upon the decomposition into truck subproblems, the complexity decreases superlinearly. Each individual truck subproblem, however, is an NP-hard problem prone to a combinatorial increase in complexity unless the problem is further decomposed into nodal subproblems. A more granular decomposition is outside the scope of the paper. Without the abovementioned decomposition, we test the limits of the subproblem-solving procedure by using off-the-shelf software. We consider the same network as before but with the addition of two nodes - 8 and 9 (Figure \ref{fig:overview}). Our findings are that while the average running time for the 5 cases in Figure \ref{Ex32} stood at 954 seconds, with
the addition of these two nodes, the running time increased significantly to 4938 seconds.

\section{Conclusion}
This paper develops a novel formulation of the joint routing and charging (JRC) problem for heavy-duty electric trucks. The optimization problem is to minimize total tardiness as well as transportation and charging costs. Through the application of decomposition and coordination principles, the proposed solution methodology is computationally efficient, allowing solving of the associated discrete optimization problem quickly. The newly developed method can obtain near-optimal solutions within a few minutes for small cases and within 30 minutes for large ones. Furthermore, it has been demonstrated that as truck battery size increases, the total cost decreases significantly; moreover, as maximum charging power increases, the number of electric trucks required decreases as well. 

\noindent \textbf{Broader Impact}. In addition to its ability to drastically reduce the complexity of the JRC problem, the proposed method also provides Lagrangian multipliers, which can be treated as ``shadow prices'' and used to provide intuitive explanations of the underlying decision-making process. The solution methodology can also run ``reactively'' and online to adapt to unexpected events such as truck/charging station breakdowns, blockage of road segments due to traffic accidents, etc. In the presence of such events, Lagrangian multipliers are readjusted ``on-the-fly'' thereby providing re-routing guidance for the trucks. Since reducing wait times upstream would reduce propagation delays downstream, a lower tardiness rate would be beneficial for supply chain and downstream operations management. 

\noindent \textbf{Future Directions}. As highlighted in \cite{baum2020modeling}, drivers may adjust speed to reduce energy consumption. Therefore, additional functionality including a choice of speed will be added to the novel formulation. Furthermore, since there may be uncertainties regarding demand \cite{chen2022integrated, kullman2021electric}, availability of charging stations \cite{guillet2022electric}, and arrival times for electric trucks \cite{froger2022electric}, the stochastic joint routing-charging problem (JRC) will also be considered. Although the above considerations add additional complexity to the solving process, preliminary results of a study in \cite{wu2023synergistic} indicate that well-trained deep neural networks can generate subproblem solutions within a matter of milliseconds.

\printnomenclature
\nomenclature[A, 01]{$bgn$}{Indicates ``begin'' for decisions when to begin charging}
\nomenclature[A, 01]{$cmplt$}{Indicates ``complete'' for decisions when to complete charging}
\nomenclature[A, 01]{$depot$}{Indicates ``depot'' for depot nodes}
\nomenclature[A, 01]{$chrg$}{Indicates ``charge'' for nodes containing chargers (if used as a superscript of a set), for charging costs and charge rates (if used as a superscript of input data) as well as for a superscript of decision variables related to charging}
\nomenclature[A, 01]{$port$}{Indicates ``port'' for port nodes}
\nomenclature[A, 01]{$trvl$}{Indicates ``travel'' for travel time}
\nomenclature[A, 01]{$dprtr$}{Indicates ``departure'' for decision variables capturing truck departure}
\nomenclature[A, 01]{$arrvl$}{Indicates ``arrival'' for decision variables capturing truck arrival}
\nomenclature[A, 01]{$trip$}{Indicates ``trip'' for decision variables capturing trips}
\nomenclature[A, 01]{$ldd$}{Indicates ``loaded'' for input data to capture the discharge rate of a truck}
\nomenclature[A, 01]{$empty$}{Indicates ``empty'' for input data to capture the discharge rate of a truck}
\nomenclature[A, 01]{$ld$}{Indicates ``load'' for binary variables that capture the ``load'' status of a truck}
\nomenclature[A, 01]{$otbnd$}{Indicates ``outbound'' for the set of trips from depot to port}
\nomenclature[A, 01]{$inbnd$}{Indicates ``inbound'' for the set of trips from port to depot}
\nomenclature[B, 01]{$N$}{Number of \underline{n}odes in the transportation network}
\nomenclature[B, 02]{$P$}{Number of time \underline{p}eriods}
\nomenclature[B, 02]{$R$}{Number of \underline{r}oad segments}
\nomenclature[B, 03]{$T$}{Number of \underline{t}rips}
\nomenclature[B, 04]{$V$}{Number of \underline{v}ehicles - trucks}
\nomenclature[B, 04]{$q^{max}$}{Candidate for the level value}
\nomenclature[B, 04]{$\beta$}{Parameter for updating penalty coefficient to penalize constraint violations}
\nomenclature[B, 04]{$\zeta$}{Parameter for updating the Polyak stepsize within SLBLR to ensure appropriate reduction of stepsizes}

\nomenclature[C, 01]{$\mathcal{N}$}{Set of \underline{n}odes}
\nomenclature[C, 02]{$\mathcal{N}^{depot}$}{Set of \underline{depot n}odes}
\nomenclature[C, 02]{$\mathcal{N}^{port}$}{Set of \underline{port n}odes: $\mathcal{N}^{depot} \cap \mathcal{N}^{port} = \emptyset$}
\nomenclature[C, 03]{$\mathcal{P}$}{Set of time \underline{p}eriods}
\nomenclature[C, 03]{$\mathcal{PR}$}{Set of \underline{pr}oducts}
\nomenclature[C, 03]{$\mathcal{R}$}{Set of \underline{r}oad segments}
\nomenclature[C, 04]{$\mathcal{T}$}{Set of \underline{t}rips}
\nomenclature[C, 05]{$\mathcal{V}$}{Set of \underline{v}ehicles - trucks}

\nomenclature[D, 01]{$n$}{\underline{N}ode number $n \in \mathcal{N} = \{1,...,N\}$}
\nomenclature[D, 02]{$n^{depot}$}{\underline{Depot n}ode number $n^{depot} \in \mathcal{N}^{depot} (\neq \emptyset) \subset \mathcal{N}$}
\nomenclature[D, 03]{$n^{port}$}{\underline{Port n}ode number $n^{port} \in \mathcal{N}^{port} (\neq \emptyset) \subset \mathcal{N}$}
\nomenclature[D, 04]{$p$}{Time \underline{p}eriod $p \in \mathcal{P} = \{1,...,P\}$}
\nomenclature[D, 05]{$r$}{\underline{R}oad segment $r \in \mathcal{R} = \{1,...,R\}$}
\nomenclature[D, 06]{$s(r)$}{\underline{S}tarting node of road segment $r$}
\nomenclature[D, 07]{$e(r)$}{\underline{E}nding node of road segment $r$}
\nomenclature[D, 08]{$t$}{\underline{T}rip number $t \in \mathcal{T} = \{1,...,T\}$}
\nomenclature[D, 09]{$v$}{\underline{V}ehicle number $v \in \mathcal{V} = \{1,...,V\}$}
\nomenclature[D, 09]{$pr_{n}$}{\underline{P}roduct that needs to be delivered to node $n$}
\nomenclature[D, 09]{$pr$}{\underline{P}roduct number}
\nomenclature[D, 09]{$k$}{Iteration number within Lagrangian Relaxation}
\nomenclature[D, 09]{$j$}{Level value update number}

\nomenclature[E, 02]{$T^{trvl}_{r,p}$}{\underline{Travel} time through \underline{r}oad segment $r$ if a truck departs from $s(r)$ at time $p$}
\nomenclature[E, 02]{$C_{p,n^{chrg}}^{chrg}$}{\underline{Charging} cost per time period if charging starts at time $p$ at node $n^{charge}$}
\nomenclature[E, 02]{$C^{lbr}$}{\underline{Labor} cost per time period after the first departure and before the last arrival (assumed to be independent of the time of the day as well as the truck)}
\nomenclature[E, 02]{$av_{v}$}{\underline{Av}ailable time of truck $v$ at the beginning of the scheduling horizon}
\nomenclature[E, 02]{$\Delta s^{dchrg,ldd}$}{\underline{Discharge} rate of a \underline{loaded} truck (in \%)}
\nomenclature[E, 02]{$\Delta s^{dchrg,empty}$}{\underline{Discharge} rate of an \underline{empty} truck (in \%)}
\nomenclature[E, 02]{$\Delta s^{chrg}$}{\underline{Charging} rate (in \%)}
\nomenclature[E, 02]{$C_{n^{chrg}}$}{Number of chargers at a charging location $n^{chrg}$}
\nomenclature[E, 01]{$due_{pr}$}{Due time for product $pr$}
\nomenclature[E, 01]{$C^{chrg}_{p,n^{chrg}}$}{\underline{Charging} cost per time period at node $n^{chrg}$ is charging starts at time $p$}

\nomenclature[F, 01]{$C^{tard}_{pr}$}{Tardiness per period penalty for late delivery of product $pr$}

\nomenclature[F, 01]{$x^{trvl,dprtr}_{v,n,p,t}$}{Binary variable that captures whether truck $v$ departed from node $n$ at time $p$ during trip $t$}
\nomenclature[F, 01]{$x^{trvl,arrvl}_{v,n,p,t}$}{Binary variable that captures whether truck $v$ arrived to node $n$ at time $p$ during trip $t$}
\nomenclature[F, 01]{$d_{v,n,t}$}{Integer variable for departure time of truck $v$ from node $n$ during trip $t$}
\nomenclature[F, 01]{$a_{v,n,t}$}{Integer variable for arrival time of truck $v$ to node $n$ during trip $t$}
\nomenclature[F, 01]{$s^{chrg}_{v,n,t}$}{State of charge continuous decision variable $\left( \in [0,1]\right)$ of a battery of truck $v$ at node $n$ during trip $t$}
\nomenclature[F, 01]{$b_{v,n^{chrg},t}$}{Integer variable for beginning of the charging time of truck $v$ at a node $n^{chrg}$ equipped with charging stations during trip $t$}
\nomenclature[F, 01]{$c_{v,n^{chrg},t}$}{Integer variable for completion of the charging time of truck $v$ at a node $n^{chrg}$ equipped with charging stations during trip $t$}
\nomenclature[F, 01]{$u_{v,n,t,pr}$}{Integer variable for product $pr$ unloading start time by truck $v$ at node $n$ (which needs to be either a port or a depot) during trip $t$}
\nomenclature[F, 01]{$x^{trip}_{v,t}$}{Binary variable to capture whether trip $t$ is taken by truck $v$}
\nomenclature[F, 01]{$x^{chrg}_{v,n^{chrg},p,t}$}{Binary variable to capture whether truck $v$ charges at node $n^{chrg}$ (quipped with chargers) at time $p$ during trip $t$}
\nomenclature[F, 01]{$\overline{u}_{pr}$}{Integer variable to capture the latest unloading time for product $pr$}
\nomenclature[F, 01]{$tard_{pr}$}{Integer variable to capture the tardiness for product $pr$}
\nomenclature[F, 01]{$\overline{a}_{v,n_{v}^{depot}}$}{Integer variables to capture the latest arrival time of vehicle $v$ at the depot $n_{v}^{depot}$}
\nomenclature[F, 01]{$\Lambda$}{Dual continuous variables (Lagrangian multipliers)}
\nomenclature[F, 01]{$sl$}{Vector of slack variables for charging station capacity constraints}
\nomenclature[F, 01]{$x^{ld}_{v,t}$}{Binary variable to capture whether truck $v$ is loaded during trip $t$}

\nomenclature[G, 01]{$O_v(d,\overline{a},x)$}{A part of the objective function for truck $v$ that includes labor and charging costs}
\nomenclature[G, 01]{$O_{pr}(tard)$}{A part of the objective function for truck $v$ that includes tardiness cost}
\nomenclature[G, 01]{$q_{\rho}(\Lambda)$}{Dual function as a function of Lagrangian multipliers $\Lambda$ and penalty coefficient $\rho$}
\nomenclature[G, 01]{$H(x)$}{Levels of constraint violations}
\nomenclature[G, 01]{$\alpha^k$}{Stepizes as a function of iteration number}
\nomenclature[G, 01]{$\rho^k$}{Penalty coefficient as a function of iteration number}
\nomenclature[G, 01]{$\overline{q}_{j}$}{A level value after $j^{th}$ update}
\nomenclature[G, 01]{$L_\rho(x,a,d,u,s,\Lambda)$}{Lagrangian function as a function of charging, departure, arrival, and trip decisions collectively denoted by $x$, arrival time $a$, departure time $d$, unloading time $u$, state of charge $s$, Lagrangian multipliers $\Lambda$ and penalty coefficient $\rho$}

\appendix
\section{Linearization of Logical Constraints} 

\subsection{Linearization of Logical Constraints with Inequalities such as \eqref{eq1}.} \label{A1}
Logical constraint \eqref{eq1} as well as other logical constraints thereafter are linearized by using big-M constraints as such:
\begin{flalign}
& d_{v,n_v^{depot},1} \geq av_v - M \cdot (1 - x^{trip}_{v,1}), \forall \left(v \in \mathcal{V}\right),   \label{eq2}
\end{flalign}
where $M$ is a large number. Indeed, if $x^{trip}_{v,1} = 1$, then $d_{v,n_v^{depot},1} \geq av_v$ as required by \eqref{eq1}; if $x^{trip}_{v,1} = 0$, then \eqref{eq1} is redundant. 

\subsection{Linearization of Logical Constraints with Equalities such as \eqref{eq6}-\eqref{eq8}.} \label{A2}
Linearization of \eqref{eq6}-\eqref{eq8} follows the same principle, with the exception that two sets of constraints are needed. For illustration purposes, linearization of \eqref{eq6} is demonstrated since other constraints are linearized in the same way: 
\begin{flalign}
& \sum_{r \in  \mathcal{R}: s(r) = n} x^{trvl,arrvl}_{v,e(r),p+T^{trvl}_{r,p}-1,t} - 1 \leq M \cdot \left(1 - x^{trvl,dprtr}_{v,n,p,t}\right), \label{eq6A1} \forall \left(v \in \mathcal{V}, n \in \mathcal{N}, t \in \mathcal{T}, p \in \mathcal{P}\right), 
\end{flalign}
\begin{flalign}
& \sum_{r \in  \mathcal{R}: s(r) = n} x^{trvl,arrvl}_{v,e(r),p+T^{trvl}_{r,p}-1,t} - 1 \geq - M \cdot \left(1 - x^{trvl,dprtr}_{v,n,p,t}\right), \label{eq6A2} \forall \left(v \in \mathcal{V}, n \in \mathcal{N}, t \in \mathcal{T}, p \in \mathcal{P}\right). 
\end{flalign}
\subsection{Linearization of Logical Constraints with Equalities and Several Logical Statements such as \eqref{eq9}-\eqref{eq11}.} \label{A3}
Linearization of \eqref{eq9} is performed as: 

\begin{flalign}
& s^{chrg}_{v,e(r),t} - \left(s^{chrg}_{v,n,t} - \frac{\Delta s^{dchrg,ldd}_{r}}{100} \cdot T^{trvl}_{r,p} \cdot x^{trvl,arrvl}_{v,e(r),p+T^{trvl}_{r,p}-1,t}\right) \leq \nonumber \\ &M \cdot \left(2 - x^{trvl,dprtr}_{v,n,p,t} - x^{trvl,arrvl}_{v,e(r),p+T^{trvl}_{r,p}-1,t}\right) + M \cdot \left(1 - x^{ld}_{v,t}\right), \label{eq9A1}
\\ & \forall \left(v \in \mathcal{V}, n \in \mathcal{N}, r \in \mathcal{R}: s(r) = n, t \in \mathcal{T}, p \in \mathcal{P}: p+T^{trvl}_{r,p}-1 \leq P\right),  \nonumber 
\end{flalign}

\begin{flalign}
& s^{chrg}_{v,e(r),t} - \left(s^{chrg}_{v,n,t} - \frac{\Delta s^{dchrg,ldd}_{r}}{100} \cdot T^{trvl}_{r,p} \cdot x^{trvl,arrvl}_{v,e(r),p+T^{trvl}_{r,p}-1,t}\right) \geq \nonumber \\ & - M \cdot \left(2 - x^{trvl,dprtr}_{v,n,p,t} - x^{trvl,arrvl}_{v,e(r),p+T^{trvl}_{r,p}-1,t}\right) - M \cdot \left(1 - x^{ld}_{v,t}\right), \label{eq9A1}
\\ & \forall \left(v \in \mathcal{V}, n \in \mathcal{N}, r \in \mathcal{R}: s(r) = n, t \in \mathcal{T}, p \in \mathcal{P}: p+T^{trvl}_{r,p}-1 \leq P\right).  \nonumber 
\end{flalign}
Accordingly, linearization of \eqref{eq10} is performed as: 
\begin{flalign}
& s^{chrg}_{v,e(r),t} - \left(s^{chrg}_{v,n,t} - \frac{\Delta s^{dchrg,ldd}_{r}}{100} \cdot T^{trvl}_{r,p} \cdot x^{trvl,arrvl}_{v,e(r),p+T^{trvl}_{r,p}-1,t}\right) \leq \nonumber \\ &M \cdot \left(2 - x^{trvl,dprtr}_{v,n,p,t} - x^{trvl,arrvl}_{v,e(r),p+T^{trvl}_{r,p}-1,t}\right) + M \cdot x^{ld}_{v,t}, \label{eq9A1}
\\ & \forall \left(v \in \mathcal{V}, n \in \mathcal{N}, r \in \mathcal{R}: s(r) = n, t \in \mathcal{T}, p \in \mathcal{P}: p+T^{trvl}_{r,p}-1 \leq P\right),  \nonumber 
\end{flalign}

\begin{flalign}
& s^{chrg}_{v,e(r),t} - \left(s^{chrg}_{v,n,t} - \frac{\Delta s^{dchrg,ldd}_{r}}{100} \cdot T^{trvl}_{r,p} \cdot x^{trvl,arrvl}_{v,e(r),p+T^{trvl}_{r,p}-1,t}\right) \geq \nonumber \\ & - M \cdot \left(2 - x^{trvl,dprtr}_{v,n,p,t} - x^{trvl,arrvl}_{v,e(r),p+T^{trvl}_{r,p}-1,t}\right) - M \cdot x^{ld}_{v,t}, \label{eq9A1}
\\ & \forall \left(v \in \mathcal{V}, n \in \mathcal{N}, r \in \mathcal{R}: s(r) = n, t \in \mathcal{T}, p \in \mathcal{P}: p+T^{trvl}_{r,p}-1 \leq P\right).  \nonumber 
\end{flalign}

Linearization of other constraints follows the same principles. For example, linearization of \eqref{eq16}, \eqref{eq21} and \eqref{eq22} follows subsection \ref{A2}, \eqref{eq17}, \eqref{eq18}, \eqref{eq19} and \eqref{eq20} follows \ref{A1} and \ref{A3}. 

\bibliographystyle{ieeetr}
\bibliography{references}

\end{document}